\documentclass[11pt,draftcls,onecolumn]{IEEEtran}
\usepackage{graphicx}
\usepackage{epsfig}
\usepackage{amsmath}
\usepackage{amssymb}
\usepackage{subfigure}
\usepackage{wrapfig}
\usepackage{times,color}
\usepackage{cite,footnote,xspace,syntonly,theorem,algorithm,algorithmic}
\input{mysymbol.sty}
\long\def\symbolfootnote[#1]#2{\begingroup
\def\thefootnote{\fnsymbol{footnote}}
\footnote[#1]{#2}\endgroup} \psfull

\begin{document}
%
\title{\textcolor{black}{Covariance Eigenvector} Sparsity for \\ Compression and Denoising}
%
%
%

\author{\emph{Ioannis~D.~Schizas and
        Georgios~B.~Giannakis$^{*}$}}

\markboth{IEEE TRANSACTIONS ON SIGNAL PROCESSING (TO APPEAR)}%
{Schizas \MakeLowercase{\textit{et al.}}: Exploiting
 Covariance-Domain Sparsity for Data Compression and Denoising}

\maketitle\maketitle \symbolfootnote[0]{$\dag$ Manuscript
received October 7, 2010; accepted January 7, 2012. First
published XXXXXX XX, 20XX; current version published XXXXXX XX,
20XX. The associate editor coordinating the review of this
manuscript and approving it for publication was Dr. Maja Bystrom. Work in this paper
was supported by the NSF Grant CCF-1016605. Part of the paper was
presented in the {\it 3rd Intl. Workshop on Comp. Advances in
Multi-Sensor Adap. Proc.}, Aruba, Dutch Antilles, Dec. 2009.}
\symbolfootnote[0]{$*$ Ioannis Schizas is with the Dept. of
Electrical Engineering, University of Texas at Arlington, 416 Yates Street, Arlington, TX 76011. Tel/fax: (817)
272-3467/272-2253. Georgios Giannakis is with the Dept. of Electrica and Computer Engineering, University of Minnesota, 200 Union Street SE, Minneapolis, MN 55455; Emails:
\texttt{\{schizas@uta.edu,georgios@umn.edu\}}.} \vspace*{-1.2cm}


\begin{abstract}
Sparsity in the \textcolor{black}{eigenvectors} of signal covariance
matrices is exploited in this paper for compression and denoising.
Dimensionality reduction (DR) and quantization modules present in
many practical compression schemes such as transform codecs, are
designed to capitalize on this form of sparsity and achieve improved
reconstruction performance compared to existing sparsity-agnostic
codecs. Using training data that may be noisy  a novel
sparsity-aware linear DR scheme is developed to fully exploit
\textcolor{black} {sparsity in the covariance eigenvectors} and form
noise-resilient estimates of the principal covariance eigenbasis.
Sparsity is effected via norm-one regularization, and the associated
minimization problems are solved using computationally efficient
coordinate descent iterations. The resulting eigenspace estimator is
shown capable of identifying a subset of the unknown support of the
eigenspace basis vectors even when the observation noise covariance
matrix is unknown, as long as the noise power is sufficiently low.
It is proved that the sparsity-aware estimator is asymptotically
normal, and the probability to correctly identify the signal
subspace basis support approaches
one, as the number of training data grows large. 
Simulations using synthetic data and
images, corroborate that the proposed algorithms achieve improved
reconstruction quality relative to alternatives.
\end{abstract}


\IEEEpeerreviewmaketitle

\newpage

\section{Introduction}\label{Sec:Intro}

Data compression has well-appreciated impact in audio, image and
video processing since the increasing data rates  cannot be
matched by the computational and storage capabilities of existing
processors. 
The cornerstone modules of compression are those performing
dimensionality reduction (DR) and quantization, as in e.g.,
transform codecs \cite{Transform_Coding}. DR projects the data
onto a space of lower dimension while minimizing an appropriate
figure of merit quantifying information loss. Quantization amounts
to digitizing the analog-amplitude DR data. Typically, DR relies
on training vectors to find parsimonious data representations with
reduced redundancy without inducing, e.g., mean-square error (MSE)
distortion in the reconstruction process. One such property that
promotes parsimony is \emph{sparsity}.

Sparsity is an attribute characterizing many natural and man-made
signals \cite{Sig_Proc_Mag}, and has been successfully exploited in
statistical inference tasks  using the least-absolute shrinkage and
selection operator (Lasso) \cite{Tibshirani_Lasso,Zou_Ada_Lasso}. In
parallel, recent results in compressive sampling rely on sparsity to
solve under-determined systems of linear equations, as well as
sample continuous signals at sub-Nyquist rates \cite{Candes_CS1}.
These tasks take advantage of sparsity present in deterministic
signal descriptors. But what if sparsity is present in statistical
descriptors such as the signal covariance matrix? The latter is
instrumental in compression when the distortion metric is MSE. In
bio-informatics and imaging applications \textcolor{black}{the data
input to the DR module have covariance matrices whose eigenvectors
admit a sparse representation over a certain domain, such as the
wavelet domain} \cite{Johnstone_SPCA}.

The `workhorse' for DR is principal component  analysis (PCA)
\cite{Brillinger_Time_Series, Jolliffe_PCA_Book}, which relies on
the covariance matrix to project the data on \textcolor{black}{the
subspace spanned by its principal eigenvectors}. So far, sparsity
has been exploited for interpreting each principal component,  but
not for reconstructing reduced-dimension signal renditions.
Specifically, the standard PCA criterion has been regularized with
an $\ell_1$-norm penalty term to induce sparsity as in the Lasso,
and perform variable selection of the data entries that
significantly contribute to each principal component
\cite{Jolliffe_Scotlass}. However, the nonconvex formulation of
\cite{Jolliffe_Scotlass} does not lend itself to efficient
optimization. The sparse PCA formulation in \cite{Zhou_Spca} leads
to a cost minimized using the block coordinate descent optimization
method \cite{Bertsekas_Nonlinear_Book}; see also
\cite{Cheng_AdaptiveSPCA} which includes a weighted $\ell_1$-norm
sparsity-imposing penalty. PCA with cardinality constraints on the
number of nonzero elements per principal eigenvector has also been
considered using relaxation techniques
\cite{Aspremont_Dspca,Aspremont_Greedy_PCA}. Alternative approaches
augment the standard singular value decomposition (SVD) cost, or the
maximum likelihood criterion with $\ell_1$ (or $\ell_0$) penalties
to effect sparsity in the principal eigenvectors \cite{Witten_Spc,
Shen_Huang_SPCA_MF, Ulfarsson_svnPCA, Ulfarsson_svnPCA_lo,
Johnstone_SPCA}. Sparsity has also been exploited to
\textcolor{black}{render PCA robust to outliers}
\cite{Candes_Robust_PCA}, as well as to reduce complexity at the
encoder \cite{Compressive_PCA}. In all the aforementioned schemes
sparsity is not exploited for reconstructing signals that have been
compressed by DR and quantization.

When dealing with  noisy data, pertinent reconstruction techniques
have been developed  to perform joint denoising and signal
reconstruction
\cite{Muresan_Adaptive_PC_Image_Denois,Simoncelli_bayesiandenoising,
Ramchadran_Image_Denois_Wav_Coefs}. However, existing approaches
rely on the noise second-order statistics being available. Here, the
a priori knowledge that the signal covariance
\textcolor{black}{eigenvectors} are sparse is exploited, and joint
denoising-reconstruction schemes are introduced without requiring
availability of the noise covariance.

The standard PCA cost is augmented with pertinent $\ell_1-$ and
$\ell_2-$norm regularization terms, that fully exploit the
sparsity present in the covariance eigenvectors when performing
not only feature extraction (as in
\cite{Jolliffe_Scotlass,Zhou_Spca,Aspremont_Dspca,Witten_Spc}) but
also reconstruction. The resulting bilinear cost is minimized via
coordinate descent which leads to an efficient sparse (S-) PCA
algorithm for DR and reconstruction.
 Its large-sample performance is analyzed in the presence of
observation noise. If the ensemble covariance matrix of the noisy
training data is known, then the sparsity-aware estimates do
better in terms of identifying the support of the principal  basis
vectors when compared to the standard sparsity-agnostic PCA. As
the number of training data used to design the DR module grows
large, the novel sparsity-aware signal covariance eigenspace
estimators: i) are asymptotically normal; and ii) identify the
true principal eigenvectors' support (indices of nonzero elements)
with probability one. The last two properties, known as `oracle'
properties \cite{Zou_Ada_Lasso}, combined with the
noise-resilience enable the novel S-PCA to attain an improved
trade-off between reconstruction performance and
reduced-dimensionality-a performance advantage also corroborated
via numerical examples. The proposed sparsity-aware DR scheme is
finally combined with a vector quantizer (VQ) to obtain a
sparsity-aware transform coder (SATC), which improves
reconstruction performance when compared to standard TC schemes
that rely on the discrete cosine transform (DCT) or PCA transform.
The merits of SATC are also demonstrated in the compression and
reconstruction/denoising of noisy images that have been extracted
from \cite{Nasa_Martian_Images}.

The rest of the paper is organized as follows. After stating the
problem setting in Sec. II, the proposed sparsity-aware PCA
formulation is introduced in Sec. III-A. Coordinate descent is
employed in Sec. III-B to minimize the associated bilinear cost,
while a computationally simpler element-wise algorithm is derived in
Sec. III-C. A cross-validation scheme is outlined in Sec. III-D for
selecting the sparsity-controlling coefficients that weigh the
$\ell_1$-norm based regularization terms. Asymptotic properties are
derived  both in the noiseless and noisy cases, establishing the
potential of the novel estimators to recover the underlying signal
covariance principal eigenvectors (Secs. IV-A, IV-B). S-PCA is
combined with vector quantization in Sec. V. Synthetic numerical
examples (Sec. VI-A) illustrate the performance advantage of SATC,
while tests using real noisy images corroborate the potential of
SATC in practical settings (Sec. VI-B).

\section{Preliminaries and Problem Formulation} \label{Sec:Prelims_Form}


Consider a collection of $n$ training data vectors
$\{\bbx_t=\bbs_t+\bbw_t\}_{t=1}^{n}$, each containing the signal
of interest $\bbs_t\in\mathbb{R}^{p\times 1}$, in additive
zero-mean possibly colored noise $\bbw_t$, assumed independent of
$\bbs_t$. It is also assumed that $\bbs_t$ lies in a linear
subspace of reduced dimension, say $r\leq p$, spanned by an
unknown orthogonal basis $\{\bbu_{s,\rho}\}_{\rho=1}^{r}$. Many
images and audio signals lie on such a low-dimensional subspace.
Accordingly, $\bbs_t$ for such signals can be expressed as
\begin{equation}\label{Eq:Signal_basis}
\bbs_t=\bbmu_{s}+\sum_{\rho=1}^{r}\pi_{t,\rho}\bbu_{s,\rho},\;\;\;t=1,\ldots,n
\end{equation}
where $\bbmu_s$ denotes the mean of $\bbs_t$, and $\pi_{t,\rho}$
are zero-mean independent projection coefficients.

The covariance matrix of the  noisy  $\bbx_t$ is given by
$\bbSigma_{x}=\bbSigma_{s}+\bbSigma_{w}$, where $\bbSigma_s$
($\bbSigma_w$) denotes the signal (noise) covariance matrix.
Consider the eigen-decomposition
$\bbSigma_{w}=\bbU_{w}\bbD_{w}\bbU_{w}^T$, where $\bbU_{w}$
denotes the eigen-matrix containing the eigenspace basis, and
$\bbD_{w}:=\textrm{diag}(d_{w,1}, \ldots d_{w,p})$ the
corresponding eigenvalues ($^T$ denotes matrix transposition).
Likewise, for the signal covariance matrix let
$\bbSigma_s=\bbU_{s}\bbD_{s}\bbU_{s}^T=\bbU_{s,r}\bbD_{s,r}\bbU_{s,r}^T$
where $\bbU_{s,r}:=[\bbu_{s,1}\ldots\bbu_{s,r}]$ and
$\bbD_{s,r}:=\textrm{diag}(d_{s,1},\ldots,d_{s,r})$ contain the
$r$ dominant eigenvectors and eigenvalues of $\bbSigma_{s}$,
respectively; while $d_{s,\rho}:={E}[\pi_{\rho,t}^2]$. Further,
let $\bbU_{s,p-r}\in\mathbb{R}^{p\times (p-r)}$ denote the matrix
formed by the subspace of dimensionality $p-r$, which is
perpendicular to the signal subspace $\bbU_{s,r}$. In the
following, $\bbmu_s$ is assumed known and subtracted from
$\bbs_{t}$; thus, without loss of generality (wlog) $\bbx_t$ and
$\bbs_t$ are assumed zero-mean. Matrices $\bbSigma_s$ and
$\bbSigma_w$ are not available, which is the case in several
applications. Moreover, the following is assumed about sample
covariances.
\newline
\noindent \textbf{(a1)} \emph{The signal of interest $\bbs_t$ and
observation noise $\bbw_t$ are independent across time $t$ and
identically distributed. Thus, by the strong law of large numbers
the sample covariance matrix estimate
$\hat{\bbSigma}_{x,n}:=n^{-1}\sum_{t=1}^{n}\bbx_t\bbx_t^T$
converges {almost surely}, as $n\rightarrow\infty$, to the
ensemble covariance matrix $\bbSigma_x$.}

\noindent Consider next a unitary transformation matrix
$\bbF\in\mathbb{R}^{p\times p}$ to form the transformed data
\begin{equation}\label{Eq:Trans_Signal}
\check{\bbx}_t:=\bbF\bbx_{t}=\sum_{\rho=1}^{r}\pi_{t,\rho}\bbF\bbu_{s,\rho}+\bbF\bbw_t.
\end{equation}
\textcolor{black}{ \noindent Such a mapping could represent the
wavelet, Fourier, or, the discrete cosine transform (DCT). The case
of interest here is when $\bbF$ is such that the transformed
eigenvectors $\check{\bbu}_{s,\rho}:=\bbF\bbu_{s,\rho}$ of
$\bbSigma_{\check{s}}$, where $\check{\bbs}:=\bbF\bbs$, have many
entries equal to zero, i.e., $\bbSigma_{\check{s}}$ has sparse
eigenvectors. \textcolor{black}{One natural question is whether
eigenvectors of a covariance matrix admit a sparse representation
over e.g., the DCT domain. Often in bio-informatics and imaging
applications the data input to the DR module have covariance matrix
with sparse eigenvectors \cite{Johnstone_SPCA}. The same attribute
is also present in other classes of signals. Consider for instance
signal vectors comprising uncorrelated groups of entries, with each
group containing correlated entries -- a case where the covariance
matrix is block diagonal. In addition to block diagonal covariance
matrices, the class also includes row- and/or column-permuted
versions of block diagonal covariance matrices.}}

\begin{figure}
\epsfig{file=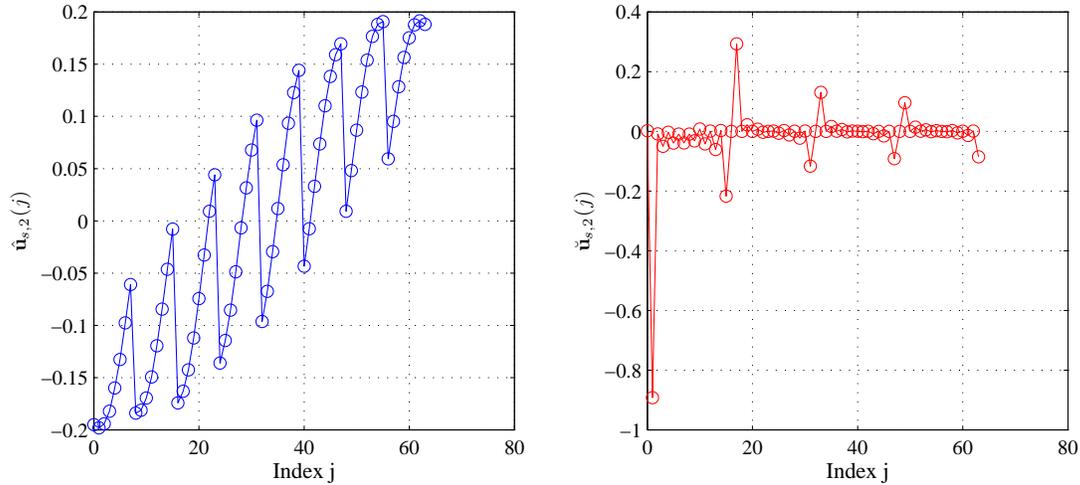,width=1\linewidth}\vspace{-0.5cm}
\caption{The second principal eigenvector of $\hat{\bbSigma}_s$
(left) and its discrete cosine transform (right).}
\label{Fig:Sparse_Second_PC}
\end{figure}

\textcolor{black}{An example is provided next to demonstrate that
$\bbSigma_{\check{s}}$ can have sparse eigenvectors. A
high-resolution image taken from \cite{Nasa_Martian_Images}
displaying part of the Martian terrain, was split into $112$ smaller
non-overlapping images of size $180\times 256$.  Each of these
images was further split into $8\times 8$ blocks. Vectors
$\{{\bbs}_t\}_{t=1}^{112}$ correspond to a block (here the $10$th
after lexicographically scanning each of the $112$ sub-images)
comprising entries with the same row and column indices in all $112$
different sub-images. An estimate of the underlying covariance
matrix of the vectorized blocks is formed using sample averaging, to
obtain
$\hat{\bbSigma}_{s}:=(112)^{-1}\sum_{t=1}^{112}\bbs_t\bbs_t^T$. The
$\textrm{rank}(\hat{\bbSigma}_s)=15$ indicates that the $64\times 1$
vectorized image blocks lie approximately in a linear subspace of
$\mathbb{R}^{64\times 1}$ having dimension $r=15$. This subspace is
spanned by the $15$ principal eigenvectors of $\hat{\bbSigma}_s$ and
forms the signal subspace. To explore whether the principal
eigenvectors of $\bbSigma_s$ have a hidden sparsity structure that
could be exploited during DR Fig. \ref{Fig:Sparse_Second_PC} (left)
depicts the value for each of the entries of the second principal
eigenvector of $\hat{\bbSigma}_s$, namely $\hat{\bbu}_{s,2}$, versus
the index number of each entry. Note that the entries of
$\hat{\bbu}_{s,2}$ exhibit a sinusoidal behavior; thus, if DCT is
applied to $\hat{\bbu}_{s,2}$ the resulting vector
$\check{\bbu}_{s,2}=\mathbf{F}\hat{\bbu}_{s,2}$ has only a few DCT
coefficients with large magnitude. Indeed, Fig.
\ref{Fig:Sparse_Second_PC} (right) corroborates that
$\hat{\bbu}_{x,2}$ is sparse over the DCT domain. In fact, all $15$
principal eigenvectors of $\hat{\bbSigma}_s$ admit a sparse
representation over the DCT domain as the one displayed in Fig.
\ref{Fig:Sparse_Second_PC} (right). Thus, the sample covariance
matrix of the transformed vectorized blocks
$\check{\bbs}_t=\bbF\bbs_t$ has $15$ principal eigenvectors that
exhibit high degree of sparsity. Such a sparse structure is to be
expected since images generated from \cite{Nasa_Martian_Images}
exhibit localized features (hilly terrain), which further result in
sparse signal basis vectors in the DCT domain\cite{Johnstone_SPCA}.
For simplicity, the original notation $\bbx_t$, $\bbs_t$ and
$\bbw_t$ will henceforth refer to the DCT transformed training data,
signal of interest, and observation noise, respectively.}

Aiming to compress data vector $\bbx$, linear DR is performed at
the encoder by left-multiplying $\mathbf{x}$ with a fat matrix
$\mathbf{C}\in\mathbf{R}^{q\times p}$, where $q \leq r\leq p$. The
reduced dimension $q$ may be chosen smaller than the signal
subspace dimension $r$, when resources are limited. Vector
$\mathbf{C}\mathbf{x}$  is received at the decoder where it is
left-multiplied by a tall $p\times q$ matrix $\bbB$ to reconstruct
$\bbs$ as $\hat{\mathbf{s}}:=\mathbf{B}\mathbf{Cx}$. Matrices
${\bbB}$ and ${\bbC}$ should be selected such that
$\hat{\mathbf{s}}=\mathbf{B}\mathbf{Cx}$ forms a `good' estimate
of $\bbs$. One pair of matrices $\mathbf{B}_o,\mathbf{C}_o$,
minimizing the reconstruction MSE
%
\begin{equation}\label{Eq:Ens_PCA}
(\bbB_o,\bbC_o)\in\arg\min_{\bbB,\bbC}
E[\|\mathbf{s}-\mathbf{BC}\mathbf{x}\|^2]
\end{equation}
are given as
$\bbC_o=\bbU_{sx,q}^T\bbSigma_{sx}\bbSigma_{x}^{-1},\;\;\bbB_o=\bbU_{sx,q}$,
where $\bbU_{sx,q}$ are the $q$ principal eigenvectors of
$\bbSigma_{sx}\bbSigma_{x}^{-1}\bbSigma_{sx}^T$, while the $\in$
notation emphasizes that \eqref{Eq:Ens_PCA} does not have a unique
optimal solution (e.g., see \cite[Ch.
10]{Brillinger_Time_Series}). In the absence of observation noise,
($\bbx_t=\bbs_t$), \eqref{Eq:Ens_PCA} corresponds to the standard
PCA, where a possible choice for $\bbB_o,\bbC_o$ is
$\mathbf{B}_{o}=\mathbf{U}_{s,q}$ and
$\mathbf{C}_{o}=\mathbf{U}_{s,q}^T$\cite[Ch.
9]{Brillinger_Time_Series}. Since the ensemble covariance matrices
are not available; $\bbC_o$ and $\bbB_o$ cannot be found. The
practical approach is to replace the cost in \eqref{Eq:Ens_PCA}
with its sample-averaged version
$n^{-1}\|\mathbf{S}-\mathbf{BC}\mathbf{X}\|_{F}^2$, where
$\bbS:=[\bbs_1\ldots\bbs_n]$ and $\bbX:=[\bbx_1\ldots\bbx_n]$.
This would require training samples for both $\bbx$, and the
signal of interest $\bbs$\cite[Ch. 10]{Brillinger_Time_Series}.
This is impossible in the noisy setting considered here.
\textcolor{black}{The reduced dimension $q\leq p$ can be selected
depending on the desired reduction viz reconstruction error
afforded.}

In the noiseless case, the optimal DR and reconstruction matrices
are formed using the signal eigenvectors, that is
$\bbC_o=\bbB_o^T=\bbU_{s,q}^T$. But even in the noisy case, the
signal subspace $\bbU_{s,q}$ is useful for joint DR and denoising
\cite{Muresan_Adaptive_PC_Image_Denois,
Ramchadran_Image_Denois_Wav_Coefs, Simoncelli_bayesiandenoising,
Yang_Past_Algorithm}. Indeed, if $\bbC=\bbB^T=\bbU_{s,q}^T$, then
it is easy to see that
$J_{\textrm{rec}}(\bbB,\bbC):=E[\|\bbs-\bbB\bbC\bbx\|_2^2]
=\textrm{tr}(\bbSigma_{w})-\textrm{tr}({\bbU}_{s,p-q}^T
\bbSigma_w{\bbU}_{s,p-q})<\textrm{tr}(\bbSigma_w)$ when $q=r<p$.
Thus, projection of $\bbx_t$ onto the signal subspace not only
achieves DR but also reduces noise effects. The question of course
is how to form an estimate for $\bbU_{s,q}$ when $\bbSigma_x$ and
$\bbSigma_w$ are unknown. Existing signal subspace estimators
assume either that i) the noise is white, namely
$\bbSigma_{w}=\sigma_{w}^2\bbI_{p}$
 for which $\bbU_{x}=\bbU_s$ ($\bbI_{p}$ denotes the $p\times p$ identity matrix);
or, ii) the $\bbSigma_w$ is known or can be estimated via
sample-averaging \cite{Muresan_Adaptive_PC_Image_Denois,
Ramchadran_Image_Denois_Wav_Coefs,
Simoncelli_bayesiandenoising,Yang_Past_Algorithm}. In the setting
here these assumptions are not needed. 

Based on training data $\{\bbx_t\}_{t=1}^{n}$, the major goal is
to exploit the sparsity present in the eigenvectors of
$\mathbf{\bbSigma}_s$ in order to derive estimates
$\hat{\bbB},\hat{\bbC}$ for the signal subspace $\bbU_{s,q}$,
thereby achieving a better trade-off between the reduced-dimension
$q$ and the MSE cost $J_{\textrm{rec}}(\bbB,\bbC)$ than existing
alternatives \cite{Jolliffe_Scotlass,Zhou_Spca,
Aspremont_Dspca,Witten_Spc}. Towards this end, a novel
sparsity-aware signal subspace estimator is developed in the next
section. Since the majority of data processing and communication
systems are digital, this sparsity-exploiting DR step will be
followed by a sparsity-cognizant VQ step in order to develop a
sparsity-aware
 transform coding (SATC) scheme for recovering $\bbs_t$ based on
quantized DR data. 

\section{Sparse Principal Component Analysis}\label{Sec:SPCA}

Recall that the standard PCA  determines DR and reconstruction
matrices $\hat{\bbC}_{o}$ and $\hat{\bbB}_o$ by minimizing the
sample-based cost
$\hat{J}_{\textrm{rec}}(\bbB,\bbC):=n^{-1}\|\bbX-\bbB\bbC\bbX\|_{F}^2$.
One possible minimizer for the latter  is
$\hat{\bbB}_o=\hat{\bbC}_o^T=\hat{\bbU}_{x,q}$, where
$\hat{\bbU}_{x,q}$ comprises the $q$ dominant eigenvectors of
$\hat{\bbSigma}_x$. In the noiseless case it holds that
$\hat{\bbSigma}_{x}=\hat{\bbSigma}_{s}$, from which it follows
that $\hat{\bbU}_{x,q}= \hat{\bbU}_{s,q}$. However, the
$q$-dominant eigenvectors of $\bbSigma_x$ do not coincide with
$\hat{\bbU}_{s,q}$ when the additive noise is colored
($\bbSigma_w\neq \sigma_w^2\bbI_{p}$). In this case, standard PCA
is not expected to estimate reliably the signal subspace.

\subsection{An $\ell_1$-regularized formulation}\label{Sec:SPCA_Est}

Here the standard PCA formulation is enhanced by exploiting the
sparsity present in $\bbU_{s,r}$. Prompted by Lasso-based sparse
regression and PCA approaches \cite{Tibshirani_Lasso,
Jolliffe_Scotlass, Zhou_Spca,Aspremont_Dspca, Witten_Spc}, the
quadratic cost of standard PCA is regularized with the $\ell_1$-norm
of the unknowns to effect sparsity. \textcolor{black}{Specifically,
$\bbB_o=\bbC_o^T=\bbU_{s,q}$ could be estimated as
\begin{equation}\label{Eq:SPCA_Nonconvex}
\arg\min_{\bbB,\bbC}n^{-1}\|\bbX-\bbB\bbC\bbX\|_{F}^2+
\sum_{\rho=1}^{q}\sum_{j=1}^{p}\lambda_{\rho}(|\mathbf{C}(\rho,j)|+|\mathbf{B}(j,\rho)|),\;\textrm{s.
to }\bbB=\bbC^T
\end{equation}
which promotes sparsity in $\bbU_{s,r}$. However, the constraint
$\bbB=\bbC^T$ leads to a nonconvex problem that cannot be solved
efficiently. This motivates the following `relaxed' version of
\eqref{Eq:SPCA_Nonconvex}, where the wanted matrices are obtained as
%
\begin{align}\label{Eq:L1_PCA}
(\hat{\bbB},\hat{\bbC})&\in\arg\min_{\bbB,\bbC}n^{-1}\|\bbX-\bbB\bbC\bbX\|_{F}^2+
\sum_{\rho=1}^{q}\sum_{j=1}^{p}\lambda_{\rho}(|\mathbf{C}(\rho,j)|+|\mathbf{B}(j,\rho)|)
+\mu\|\bbB-\bbC^T\|_{F}^2
\end{align}
using efficient coordinate descent solvers.} Note that $q\leq r$,
since the dimensionality of the signal subspace may not be known.
Moreover, $\{\lambda_{\rho}\}_{\rho=1}^q$ are nonnegative constants
controlling the sparsity of $\mathbf{B}$ and $\mathbf{C}$. Indeed,
the larger $\lambda_{\rho}$'s are chosen, the closer the entries of
$\bbB$ and $\bbC$ are driven to the origin. \textcolor{black}{Taking
into account that the `clairvoyant' compression and reconstruction
matrices satisfy $\bbB_o=\bbC_o^T=\bbU_{s,q}$, the term
$\mu\|\bbB-\bbC^T\|_F^2$ ensures that $\hat{\bbB}$ and
$\hat{\bbC}^T$ stay close.} \textcolor{black}{Although $\bbB_o$ and
$\bbC_o$ are orthonormal, $\bbB$ and $\bbC$ are not constrained to
be orthonormal in \eqref{Eq:L1_PCA} because orthonormality
constraints of the form $\bbB^T\bbB=\mathbf{I}$ and
$\bbC\bbC^T=\mathbf{I}$ are nonconvex. With such constraints
present, efficient coordinate descent algorithms converging to a
stationary point cannot be guaranteed.}

\textbf{Remark 1:} The minimization problem in \eqref{Eq:L1_PCA}
resembles the sparse PCA formulation proposed in \cite{Zhou_Spca}.
However, the approach followed in \cite{Zhou_Spca} imposes sparsity
\emph{only} on $\mathbf{C}$, while it forces matrix $\mathbf{B}$ to
be orthonormal. The latter constraint mitigates scaling issues
(otherwise $\mathbf{C}$ could be made arbitrarily small by
counter-scaling $\mathbf{B}$), but is otherwise not necessarily
well-motivated. Without effecting sparsity in $\bbB$, the
formulation in \cite{Zhou_Spca} does not fully exploit the sparsity
present in the eigenvectors of $\bbSigma_{s}$, which further results
in sparse clairvoyant matrices $\mathbf{C}_o$ and $\mathbf{B}_o$ in
the absence of noise. Notwithstanding, \eqref{Eq:L1_PCA} combines
the reconstruction error $n^{-1}\|\mathbf{X}-\mathbf{BCX}\|_F^2$
with regularization terms that impose sparsity to \emph{both}
$\mathbf{B}$ and $\mathbf{C}$. \textcolor{black}{Even though the
$\ell_1$-norm of $\bbC$ together with $\|\bbB-\bbC^T\|_F^2$ suffice
to prevent scaling issues, the $\ell_1$-norm of $\bbB$ is still
needed to ensure sparsity in the entries of $\hat{\bbB}$.}

\subsection{Block Coordinate Descent Algorithm}\label{Sec:SPCA_BCD}

The minimization problem  in \eqref{Eq:L1_PCA} is nonconvex with
respect to (wrt) both $\mathbf{B}$ and $\mathbf{C}$. This
challenge will be bypassed by constructing an iterative solver.
Relying on block coordinate descent (see e.g., \cite[pg.
160]{Bertsekas_Nonlinear_Book}) the cost in \eqref{Eq:L1_PCA} will
be iteratively minimized wrt $\mathbf{B}$ (or $\mathbf{C}$), while
keeping matrix $\bbC$ (or $\bbB$) fixed.

Specifically, given the matrix $\hat{\bbB}_{\tau-1}$ at the end of
iteration $\tau-1$, an updated estimate of $\bbC_o$ at iteration
$\tau$ can be formed by solving the minimization problem [cost in
\eqref{Eq:L1_PCA} has been scaled with $n$]
\begin{equation}\label{Eq:C_tauplus1}
\hat{\bbC}_{\tau}=\arg\min_{\bbC}\|\bbX-\hat{\bbB}_{\tau-1}
\bbC\bbX\|_{F}^{2}+\sum_{\rho=1}^{q}\lambda_{\rho}\|\bbC_{\rho:}^T\|_{1}
+\mu\|\bbC-\hat{\bbB}_{\tau-1}^T\|_{F}^{2}
\end{equation}
where $\bbC_{\rho:}^T$ denotes the $\rho$th row of $\bbC$,
\textcolor{black}{while $n$ is absorbed in $\mu$ and
$\lambda_{\rho}$}. After straightforward manipulations
\eqref{Eq:C_tauplus1} can be equivalently reformulated as
\begin{equation}\label{Eq:C_tauplus1_b}
\hspace{-0.2cm}\hat{\bbC}_{\tau}=\arg\min_{\bbC}
\textrm{tr}(\bbX^T\bbC^T\hat{\bbB}_{\tau-1}^T\hat{\bbB}_{\tau-1}\bbC\bbX
-2\bbX^T\bbC^T\hat{\bbB}_{\tau-1}^T\bbX)+
\sum_{\rho=1}^{q}[\lambda_{\rho}\|\bbC_{\rho:}^T\|_{1}
+\mu\|\bbC_{\rho:}^T\|_2^2-2\mu\bbC_{\rho:}^T\hat{\bbB}_{\tau-1,:\rho}]
\end{equation}
where $\hat{\bbB}_{\tau-1,:\rho}$ corresponds to the $\rho$th
column of $\hat{\bbB}_{\tau-1}$. After introducing some auxiliary
variables $\{t_{\rho,j}\}_{j=1,\rho=1}^{p,q}$, the optimization
problem in \eqref{Eq:C_tauplus1_b} can be equivalently rewritten
as a convex optimization problem that has i) a cost given by
$\textrm{tr}(\bbX^T\bbC^T\hat{\bbB}_{\tau-1}^T\hat{\bbB}_{\tau-1}$
$\bbC\bbX)-2\textrm{tr}(\bbX^T\bbC^T\hat{\bbB}_{\tau-1}^T\bbX)+
\sum_{\rho=1}^{q}\lambda_{\rho}\sum_{j=1}^{p}t_{\rho,j}+\mu\sum_{\rho=1}^{q}
\|\bbC_{\rho:}^T\|_2^2-2\mu\sum_{\rho=1}^{q}\bbC_{\rho:}^T\hat{\bbB}_{\tau-1,:\rho}$;
and ii) a constraint set formed by the inequalities
$\{|\bbC(\rho,j)|\leq t_{\rho,j}\}_{j=1,\rho=1}^{p,q}$. This
constrained minimization problem can be solved using an interior
point method \cite{Boyds_Book}.

Given the most recent DR  update $\hat{\bbC}_{\tau}$, an updated
estimate of the reconstruction matrix ${\bbB}_{o}$ is obtained as
\begin{equation}\label{Eq:Btaup1}
\hat{\bbB}_{\tau}=\arg\min_{\bbB}
\|\bbX-\bbB\hat{\bbC}_{\tau}\bbX\|_{F}^2+\sum_{\rho=1}^{q}
\lambda_{\rho}\|\bbB_{:\rho}\|_{1}+
\mu\|\bbB-\hat{\bbC}_{\tau}^T\|_{F}^2.
\end{equation}
The minimization problem in \eqref{Eq:Btaup1} can be split into
the following $p$ subproblems:
\begin{align}\label{Eq:Btaup1_row_l}
\hat{\bbB}_{\tau,j:} &=\arg\min_{\bbB_{j:}}
\|\bbX_{j:}-\bbB_{j:}^T\hat{\bbC}_{\tau}\bbX\|_2^2+\mu\|\bbB_{j:}^T-
(\hat{\bbC}_{\tau,:j})^T\|_{2}^2
+\sum_{\rho=1}^{q}\lambda_{\rho}|\bbB(j,\rho)|\nonumber\\
&=\arg\min_{\bbB_{j:}}\|[\bbX_{j:}\;,\;\mu^{1/2}(\hat{\bbC}_{\tau,:j})^T]
-\bbB_{j:}^T[\hat{\bbC}_{\tau}\bbX\;,\;\mu^{1/2}\mathbf{I}_{q}]\|_2^2+
\sum_{\rho=1}^{q}\lambda_{\rho}|\bbB(j,\rho)|,\;\;j=1,\ldots,p
\end{align}
where $\bbX_{j:}$ denotes the $j$th row of $\bbX$.

Notice that \eqref{Eq:Btaup1_row_l} corresponds  to a Lasso
problem that can be solved efficiently using e.g., the LARS
algorithm \cite{Efron_Hastie_LARS}. The proposed block coordinate
descent (BCD-) S-PCA algorithm yields iterates $\hat{\bbB}_{\tau}$
and $\hat{\bbC}_{\tau}$ that converge, as $\tau\rightarrow\infty$,
at least to a stationary point of the cost in \eqref{Eq:L1_PCA}- a
fact established using the results in e.g.,
\textcolor{black}{\cite[Thm. 4.1]{BCD_Tseng} (see also arguments in
Apdx. B)}. The BCD-SPCA scheme is tabulated as Algorithm
\ref{Alg:BCD_SPCA}.
\begin{algorithm}[t]
\caption{: BCD-SPCA} \small{
\begin{algorithmic}
    \STATE Initialize $\hat{\bbB}_{0}=\hat{\bbC}_0^T=\hat{\bbU}_{s,q}$, where
    $\hat{\bbU}_{s,q}$ is the signal subspace estimate found by standard PCA.
    \FOR {$\tau=1$,$\ldots$}

        \STATE Find $\hat{\bbC}_{\tau}$ by solving \eqref{Eq:C_tauplus1_b}.

        \STATE Find $\hat{\bbB}_{\tau}$ by solving \eqref{Eq:Btaup1_row_l}
         for $j=1,\ldots,p$.

        \STATE {\textbf{If} $|\textrm{Cost}_{\tau}-\textrm{Cost}_{\tau-1}|<
        \epsilon$ for a prescribed tolerance $\epsilon$ \textbf{then} break}
    \ENDFOR
\end{algorithmic}}
\label{Alg:BCD_SPCA}
\end{algorithm}


\subsection{Efficient SPCA Solver}\label{Sec:SPCA_eff}
Relying on the BCD-SPCA algorithm of the previous section, an
element-wise coordinate descent algorithm is developed here to
numerically solve \eqref{Eq:L1_PCA} with reduced computational
complexity. Specifically, the cost in \eqref{Eq:L1_PCA} is
iteratively minimized wrt an entry of either $\mathbf{B}$ or
$\mathbf{C}$, while keeping the remaining entries fixed. One
coordinate descent iteration involves updating all the entries of
matrices $\mathbf{B}$ and $\mathbf{C}$.

Given iterates $\hat{\bbB}_{\tau-1}$ and $\hat{\bbC}_{\tau-1}$,
the next steps describe how the entries of $\hat{\bbC}_{\tau}$ and
$\hat{\bbB}_{\tau}$ are formed. Let $\otimes$ denote the Kronecker
product, and $\textrm{vec}(\bbC)$ the $qp\times 1$ vector obtained
after stacking the $p$ columns of $\mathbf{\bbC}$. Using the
property
$\textrm{vec}({\hat{\bbB}_{\tau-1}\bbC\bbX})=(\mathbf{X}^T\otimes
\hat{\mathbf{B}}_{\tau-1})\textrm{vec}(\bbC)$, the cost in
\eqref{Eq:L1_PCA} after setting $\bbB=\hat{\bbB}_{\tau-1}$ can be
re-expressed as
\begin{equation}\label{Eq:MSE_S1_Kronecker}
\|\textrm{vec}({\bbX})-({\bbX}^T\otimes{\hat{\bbB}}_{\tau-1})\textrm{vec}({\bbC})\|_{2}^2
+\sum_{\rho=1}^{q}\lambda_{\rho}(\|\mathbf{C}_{\rho:}^T\|_{1}+
\|\hat{\mathbf{B}}_{\tau-1,:\rho}\|_{1})
+\mu\|\hat{\bbB}_{\tau-1}-\bbC^T\|_{F}^2.
\end{equation}
Next, we show how to form $\hat{\bbC}_{\tau}(\rho,j)$ at iteration
$\tau$, based on the most up-to-date values of $\mathbf{B}$ and
$\bbc_{v}:=\textrm{vec}(\bbC)$, namely
$\hat{\mathbf{B}}_{\tau-1}$,
$\{\hat{\mathbf{c}}_{v,\tau-1}(m)\}_{m=(j-1)q+\rho+1}^{qp}$ and
$\{\hat{\mathbf{c}}_{v,\tau}(m)\}_{m=1}^{(j-1)q+\rho-1}$. It follows from
\eqref{Eq:MSE_S1_Kronecker} that
$\hat{\bbC}_{\tau}(\rho,j)\equiv\hat{\mathbf{c}}_{v,\tau}((j-1)q+\rho)$,
for $\rho=1,\ldots,q$ and $j=1,\ldots,p$, can be determined as
\begin{equation}\label{Eq:Elementwise_minimization}
\hat{\bbC}_{\tau}(\rho,j)=\arg\min_{c}\|\bbchi_{\tau,\rho,j}
-c\hat{\bbX}_{B_{\tau-1},:(j-1)q+\rho}\|_2^2
+\mu(c-\hat{\bbB}_{\tau-1}(j,\rho))^2+\lambda_{\rho}|c|
\end{equation}
where $\hat{\bbX}_{B_{\tau-1}}:=\bbX^T\otimes\hat{\bbB}_{\tau-1}$,
$\bbchi_{\tau,\rho,j}:=\textrm{vec}(\bbX)-\textstyle\sum_{m=1}^{(j-1)q+\rho-1}
\hat{\bbc}_{v,\tau}(m)\hat{\bbX}_{B_{\tau-1},:m}
-\textstyle\sum_{m=(j-1)q+\rho+1}^{qp} \hat{\bbc}_{v,\tau-1}(m)$
$\hat{\bbX}_{B_{\tau-1},:m}$, and $\hat{\bbX}_{B_{\tau-1},:m}$
corresponds to the $m$th column of $\hat{\bbX}_{B_{\tau-1}}$.
Interestingly, the minimization in
\eqref{Eq:Elementwise_minimization} corresponds to a sparse
regression (Lasso) problem involving a scalar. The latter admits a
closed-form solution which is given in the next Lemma (see Apdx. A
for the proof).
\begin{lemma}\label{Lem:Scalar_Lasso}
The optimal solution of the minimization problem
\begin{equation}\label{Eq:Scalar_Lasso}
\hat{c}=\arg\min_{c}\|\bbchi-c\bbh\|_2^2+\mu(c-\hat{b})^2+\lambda|c|
\end{equation}
where $\bby$ and $\bbh$ are column vectors and $\hat{b}$
is a scalar constant, is given by
$$\hat{c}=\textrm{\emph{sgn}}\left({\bbchi^T\bbh+\mu\hat{b}}\right)
\times\left(\left|\frac{\bbchi^T\bbh+\mu\hat{b}}{\|\bbh\|_2^2+\mu}\right|
-\frac{\lambda}{2\|\bbh\|_2^2+2\mu}\right)_{+}.$$
\end{lemma}
Based on Lemma \ref{Lem:Scalar_Lasso} one can readily deduce that
the optimal solution of \eqref{Eq:Elementwise_minimization} is
given by
\begin{align}\label{Eq:C_tau_i_j}
\hat{\bbC}_{\tau}(\rho,j)=&
\textrm{sgn}\left({(\bbchi_{\tau,\rho,j})^T\hat{\bbX}_{B_{\tau-1},:(j-1)q+\rho}
+\mu\hat{\bbB}_{\tau-1}(j,\rho)}\right)\nonumber\\
&\times\left(\left|\frac{(\bbchi_{\tau,\rho,j})^T\hat{\bbX}_{B_{\tau-1},:(j-1)q+\rho}
+\mu\hat{\bbB}_{\tau-1}(j,\rho)}
{\|\hat{\bbX}_{B_{\tau-1},:(j-1)q+\rho}\|_2^2+\mu}\right|
-\frac{\lambda_{\rho}}{2\|\hat{\bbX}_{B_{\tau-1},:(j-1)q+\rho}\|_2^2+2\mu}\right)_{+}
\end{align}
where $(\cdot)_{+}:=\max(\cdot,0)$.

Similarly, starting from the minimization problem in \eqref{Eq:Btaup1_row_l}
and applying an element-wise coordinate descent approach
an update for the $\bbB(j,\rho)$ can be obtained as
\begin{equation}\label{Eq:Btau_row_l_col_j}
\hat{\bbB}_{\tau}(j,\rho)=
\arg\min_{\bbb}\|\bbpsi_{\tau,j,\rho}-b\hat{\bbX}_{C_{\tau},:\rho}\|_2^2
+\mu(b-\hat{\bbC}_{\tau}(\rho,j))^2+\lambda_{\rho}|b|,\;\;\rho=1,\ldots,q
\end{equation}
where $\hat{\bbX}_{C_{\tau}}:=\bbX^T\hat{\bbC}_{\tau}^T$,
$\bbpsi_{\tau,j,\rho}:=\bbX_{:j}-\textstyle\sum_{l=1}^{\rho-1}
\hat{\bbB}_{\tau}(j,l)\hat{\bbX}_{C_{\tau},:l}
-\textstyle\sum_{l=\rho+1}^{q}\hat{\bbB}_{\tau-1}(j,l)
\hat{\bbX}_{C_{\tau},:l}$, and $\hat{\bbX}_{C_{\tau},:l}$ denotes
the $l$th column of $\hat{\bbX}_{C_{\tau}}$, while $\bbX_{:j}$
refers to the $j$th column of $\bbX$. The optimal solution of the
minimization problem in \eqref{Eq:Btau_row_l_col_j} is given by
\begin{equation}\label{Eq:B_tau_j_rho}
\hat{\bbB}_{\tau}(j,\rho)=\textrm{sgn}
\left({(\bbpsi_{\tau,j,\rho})^T\hat{\bbX}_{C_{\tau},:\rho}+
\mu\hat{\bbC}_{\tau}(\rho,j)}\right)\times \left(
\left|\frac{(\bbpsi_{\tau,j,\rho})^T\hat{\bbX}_{C_{\tau},:\rho}
+\mu\hat{\bbC}_{\tau}(\rho,j)}{\|
\hat{\bbX}_{C_{\tau},:\rho}\|^2_2+\mu}\right|-\frac{\lambda_{\rho}}{2\|
\hat{\mathbf{X}}_{C_{\tau},:\rho}\|_2^2+2\mu}
\right)_{+}.\hspace{-0.5cm}
\end{equation}
Note that iteration $\tau$ involves minimizing \eqref{Eq:L1_PCA} wrt
to each entry of $\mathbf{C}$ or $\mathbf{B}$ while fixing the rest.
\textcolor{black}{It is shown in Appendix B that the computationally
efficient coordinate descent (ECD)-SPCA scheme converges, as
$\tau\rightarrow\infty$, at least to a stationary point of the cost
in \eqref{Eq:L1_PCA} when the entries of $\bbX$ are finite. The
proof relies on \cite[Thm. 4.1]{BCD_Tseng}. Using arguments similar
to those in Appendix B, it can be shown that BCD-SPCA converges too.
A stationary point for the nondifferentiable cost here is defined as
one whose lower directional derivative is nonnegative toward any
possible direction \cite[Sections 1 and 3]{BCD_Tseng}, meaning that
the cost cannot \emph{decrease} when moving along any possible
direction around and close to a stationary point. A strictly
positive $\mu$ ensures that the minimization problems in
\eqref{Eq:Elementwise_minimization} and \eqref{Eq:Btau_row_l_col_j}
are \emph{strictly} convex with respect to either $\bbC$ or $\bbB$.
This guarantees that the minimization problems in
\eqref{Eq:Elementwise_minimization} or \eqref{Eq:Btau_row_l_col_j}
have a unique minimum per iteration, which in turn implies that the
ECD-SPCA algorithm converges to a stationary point. If $\mu=0$, the
proposed algorithms may not converge. This can also be seen from the
updating recursions \eqref{Eq:C_tau_i_j} and \eqref{Eq:B_tau_j_rho}.
If $\mu=0$ and at a certain iteration one of the matrices is zero,
say $\hat{\bbC}_{\tau}$, this may cause the entries of
$\hat{\bbB}_{\tau}$ to diverge. Similar comments apply in BCD-SPCA.}

The optimal solution in \eqref{Eq:C_tau_i_j} and
\eqref{Eq:B_tau_j_rho} can be determined in closed form at
computational complexity  ${\cal O}(np)$. With $2qp$ entries in
$\bbC$ and $\bbB$, the total complexity for completing a full
coordinate descent iteration is ${\cal O}(nqp^2)$. The ECD-SPCA
scheme is tabulated as Algorithm \ref{Alg:ECD_SPCA}.
\begin{algorithm}[t]
\caption{: ECD-SPCA} \small{
\begin{algorithmic}
    \STATE Initialize $\hat{\bbB}_{0}=\hat{\bbC}_0^T=\hat{\bbU}_{s,q}$, where
    $\hat{\bbU}_{s,q}$ is the signal subspace estimate found by standard PCA.
    \FOR {$\tau=1$,$\ldots$}

        \STATE For $j=1,\ldots,p$ and
            $\rho=1,\ldots,q$ determine $\hat{\bbC}_{\tau}(\rho,j)$
            using
            \eqref{Eq:C_tau_i_j}.

        \STATE For $j=1,\ldots,p$ and
            $\rho=1,\ldots,q$ determine $\hat{\bbB}_{\tau}(j,\rho)$
            using
            \eqref{Eq:B_tau_j_rho}.

        \STATE {\textbf{If} $|\textrm{Cost}_{\tau}-\textrm{Cost}_{\tau-1}|<
        \epsilon$ for a prescribed tolerance $\epsilon$ \textbf{then} break}
    \ENDFOR
\end{algorithmic}}
\label{Alg:ECD_SPCA}
\end{algorithm}

Note that the sparsity coefficient $\lambda_{\rho}$ is  common to
both  terms $\|\bbC_{\rho:}^T\|_{1}$ and $\|\bbB_{:\rho}\|_{1}$.
This together with the explicit dissimilarity penalty in
\eqref{Eq:L1_PCA} force the estimates obtained via ECD-SPCA (denoted
$\hat{\mathbf{B}}_{\tau}$ and $\hat{\mathbf{C}}_{\tau}^T$) to be
approximately equal for sufficiently large $\tau$. The latter
equality requirement is further enforced in the BCD- (or ECD-) SPCA
scheme by selecting $\mu$ sufficiently large, e.g., in our setting
$\mu=100$. The ideal $\bbB_o$ and $\bbC_o^T$ are orthonormal and
equal in the noiseless case; thus, the same properties are imposed
to $\hat{\mathbf{B}}_{\tau}$ and $\hat{\mathbf{C}}_{\tau}^T$.
Towards this end, we: i) pick one of the matrices
$\hat{\mathbf{B}}_{\tau}$ and $\hat{\mathbf{C}}_{\tau}^T$, say the
latter; ii) extract, using SVD, an orthonormal basis
$\hat{\hat{\mathbf{U}}}_{s}\in\mathbb{R}^{p\times q}$  spanning the
range space of $\hat{\mathbf{C}}_{\tau}^T$; and iii) form the
compression and reconstruction matrices
$\hat{\hat{\mathbf{B}}}_{\tau}=\hat{\hat{\mathbf{C}}}_{\tau}^T
=\hat{\hat{\mathbf{U}}}_{s}$. Thus, the dimensionality of each
acquired vector $\bbx$ is reduced at the encoder using the linear
operator $\hat{\hat{\mathbf{U}}}_{s}$, while at the decoder the
signal of interest is reconstructed as
$\hat{\bbs}=\hat{\hat{\bbU}}_{s}(\hat{\hat{\bbU}}_{s}^T\bbx)=
(\hat{\bbC}_{\tau}^T\hat{\bbC}_{\tau})^{\dag}\hat{\bbC}_{\tau}^T
(\hat{\bbC}_{\tau}\bbx)$, where the symbol $\dag$ denotes matrix
pseudoinverse.

\subsection{Tuning the sparsity-controlling coefficients}
\label{Sec:Setting_lambda} Up to now the sparsity-controlling
coefficients were assumed given. A cross-validation (CV) based
algorithm is derived in this section to select
$\{\lambda_{\rho}\}_{\rho=1}^q$ with the objective to find
estimates $\hat{\bbB}$ and $\hat{\bbC}$ leading to good
reconstruction performance, i.e., small
$J_{\textrm{rec}}(\hat{\bbB},\hat{\bbC})$. The CV scheme is
developed for the noiseless case ($\bbx_t=\bbs_t$).

Consider the $M$-fold CV scheme   in \cite[pg.
241-249]{Hastie_Statistical_Learning}, for selecting
$\{\lambda_{\rho}\}_{\rho=1}^{q}$ from a $q$-dimensional grid
${\cal L}:=\{\lambda_1^1,\ldots,\lambda_1^{J}\}
\times\ldots\times\{\lambda_{q}^{1},\ldots,\lambda_{q}^{J}\}$,
where
$\{0<\lambda_{\rho}^1<\ldots<\lambda_{\rho}^{J}\}_{\rho=1}^{q}$
denote the candidate values, and $\times$ denotes Cartesian
product. The training data  set $\{\bbx_t\}_{t=1}^{n}$ is divided
into $M$ nonoverlapping subsets $\{\bbX_{m}\}_{m=1}^{M}$. Let
$\hat{\bbB}_{-m}(\{\lambda_{\rho}\}_{\rho=1}^{q})$ and
$\hat{\bbC}_{-m}(\{\lambda_{\rho}\}_{\rho=1}^{q})$ denote the
estimates obtained via BCD-SPCA (or ECD-SPCA) when using all the
training data but those in $\bbX_{m}$, for fixed values of the
sparsity-controlling coefficients. The next step is to evaluate an
estimate of the reconstruction error using $\bbX_{m}$, i.e., form
the reconstruction error estimate
$\hat{J}_{\textrm{rec}}^m(\{\lambda_{\rho}\}_{\rho=1}^{q}):
=|\bbX_{m}|^{-1}\|\bbX_{m}-\hat{\bbB}_{-m}\hat{\bbC}_{-m}\bbX_{m}\|_{F}^2$,
where $|\bbX_{m}|$ indicates the cardinality of $\bbX_{m}$. A
sample-based estimate of the reconstruction MSE can be found as
\begin{equation}\label{Eq:CV_Estimate}
\hat{J}_{\textrm{rec}}(\{\lambda_{\rho}\}_{\rho=1}^{q})
=n^{-1}\sum_{t=1}^{n}\hat{J}_{\textrm{rec}}^{m(t)}(\{\lambda_{\rho}\}_{\rho=1}^{q})
\end{equation}
where $m(t)$ denotes the partition index in which $\bbx_t$ is
included during the CV process.

\begin{figure}
\centerline{\epsfig{file=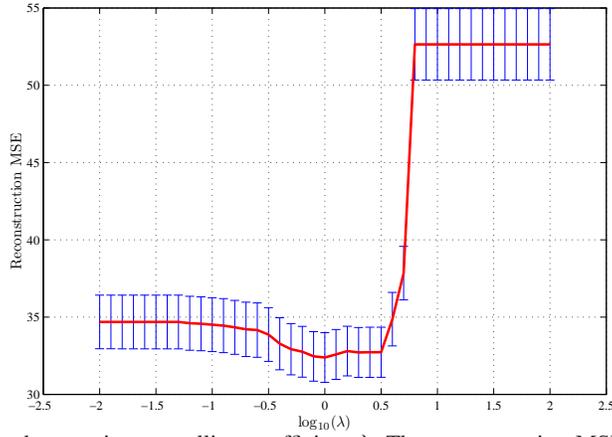,width=3.8in}}\vspace{-1.0cm}
\caption{Reconstruction MSE vs. the sparsity-controlling
coefficient $\lambda$. The reconstruction MSE is estimated via
$5$-fold CV.} \label{Fig:Sparse_Eigenspace}
\end{figure}

Using \eqref{Eq:CV_Estimate}, the desired sparsity-controlling
coefficients are selected as
\begin{equation}\label{Eq:Opt_Lambdas}
(\{\hat{\lambda}_{\rho}\}_{\rho=1}^{q}):=\arg\min_{\{\lambda_{\rho}\}\in {\cal L}}
\hat{J}_{\textrm{rec}}(\{\lambda_{\rho}\}_{\rho=1}^{q}).
\end{equation}
The minimization \eqref{Eq:Opt_Lambdas} is carried out using
exhaustive search over the grid ${\cal L}$. Fig.
\ref{Fig:Sparse_Eigenspace} (right) shows how the reconstruction
error $\hat{J}_{\textrm{rec}}(\cdot)$ is affected by the sparsity
controlling coefficients. A simplified scenario is considered here
with $\{\lambda_{\rho}=\lambda\}_{\rho=1}^q$, with $p=14$, and
$q=r=2$. Matrix $\bbSigma_{x}=\bbSigma_{s}$ is constructed so that
$80\%$ of the  $\bbU_s$ entries are zero. The black bars in Fig.
\ref{Fig:Sparse_Eigenspace} (right) quantify the standard error
associated with each reconstruction MSE estimate in the red CV
curve and its amplitude is calculated as
\begin{equation}\label{Eq:Standard_Error}
\textrm{Std}(\lambda)=\sqrt{n^{-1}}\sqrt{(n-1)^{-1}
\sum_{t=1}^{n}[\hat{J}_{\textrm{rec}}^{m(t)}(\lambda)-\hat{J}_{\textrm{rec}}(\lambda)]^2}.
\end{equation}
When $\lambda<0.1$ the reconstruction MSE remains almost constant
and equal to the one achieved by standard PCA ($\lambda=0$). If
$\lambda>10^{0.5}$, the reconstruction MSE increases and reaches a
maximum equal to the trace of $\bbSigma_x$ (the DR and
reconstruction matrices equal zero). The minimum MSE is achieved
for $\lambda\approx 1$. Note that $\lambda_{\rho}$'s have so far
been selected for a fixed value of $\mu$. Recall that $\mu$
controls the dissimilarity, of $\hat{\bbC}^T$ with $\hat{\bbB}$,
thus a relatively large value  ($\mu=100$ was used in the
simulations)  suffices to ensure that  $\hat{\bbC}$ and
$\hat{\bbB}$ stay close. Of course, the higher $\mu$ is the closer
$\hat{\bbC}^T$ and $\hat{\bbB}$ will be.

\section{S-PCA Properties }\label{Sec:Properties_SPCA}
In this section sufficiently many training data are assumed
available ($n\rightarrow\infty$), to allow analysis based on the
ensemble covariance $\bbSigma_x=\bbSigma_{s}+\bbSigma_w$. Recall
that (a1) ensures a.s. convergence of the sample-based cost in
\eqref{Eq:L1_PCA} to its ensemble counterpart
\begin{equation}\label{Eq:Ens_L1_PCA}
({\bbB}_e,{\bbC}_e)\in\arg\min_{\bbB,\bbC}
\textrm{tr}[(\bbI-\bbB\bbC)\bbSigma_{x}(\bbI-\bbB\bbC)^T]
+\sum_{\rho=1}^{q}\lambda_{\rho}(\|\bbC_{\rho:}^T\|_1+\|\bbB_{:\rho}\|_1)
+\mu\|\bbB-\bbC^T\|_F^2.
\end{equation}
%
%
Interestingly, it will turn out that even in the presence of colored
noise $\bbw_t$, the solution pair ${\bbB}_{e},{\bbC}_{e}$ can
recover the support of the columns of the signal subspace
$\bbU_{s,r}$, or at least part of it, as long as the noise power in
$\bbx_t$ is sufficiently small.

\textcolor{black}{\subsection{Support recovery in colored noise}
}\label{Sec:Noiseless}
In this section entries of $\bbC_{e}$ (or $\bbB_e$) will be
considered nonzero only if their corresponding magnitudes exceed
an arbitrarily small threshold $\delta>0$. Under this condition it
will be demonstrated next that for properly selected $\mu$ and
$\lambda_{\rho}$, S-PCA assigns the (non)zero entries in
$\bbC_{e}$ consistently with the support of the columns of
$\bbU_{s,r}$. This means that the S-PCA formulation is meaningful
because it does not assign entries of $\bbC_{e}$ (or $\bbB_e$)
arbitrarily, but takes into account where the (non)zero entries of
$\bbU_{s,r}$ are. Interestingly, this will hold even when colored
noise is present, as long as its variance is properly upper
bounded.

To proceed, let $\bbSigma_{s}$ be a permuted version of a block
diagonal matrix. Specifically:
\newline
\textbf{(a2)} \emph{The entries of  $\bbs$ can be partitioned into
groups ${\cal G}_1,\ldots,{\cal G}_{K}$, so that entries with
indices in the same group are allowed to be correlated but entries
with indices in different groups are uncorrelated; i.e., if $j \in
{\cal G}_k$ and $j' \in {\cal G}_{k'}$, then
$E[\mathbf{s}(j)\mathbf{s}(j')]=0$ for $k\neq k'$. Moreover, let
$\{{\cal G}_{k}\}_{k=1}^{K}$ have the same cardinality that is
equal to $G$. Using the proper permutation matrix $\mathbf{P}$,
these groups can be made contiguous; hence, the vector
$\mathbf{s}_{P}: =\mathbf{Ps}=[\mathbf{s}_{{\cal
G}_1}\ldots\mathbf{s}_{{\cal G}_K}]^T$ has covariance matrix with
block diagonal structure, that is
$\mathbf{P}\mathbf{\Sigma}_s\mathbf{P}^T=
\textrm{\emph{bdiag}}(\mathbf{\Sigma}_{{s}_{{\cal
G}_1}},\ldots,\mathbf{\Sigma}_{{s}_{{\cal G}_K}})$. This implies
that the eigenvector matrix $\tilde{\mathbf{U}}_{s}$ of
$\mathbf{P}\mathbf{\Sigma}_s\mathbf{P}^T$ is block diagonal and
sparse. Since $\mathbf{U}_{s}=\mathbf{P}^T\tilde{\mathbf{U}}_{s}$
and $\bbP$ is a permutation matrix,  it follows that
$\mathbf{U}_{s}$ is also sparse.}
\newline
The block-diagonal structure under (a2) emerges when $\bbs_t$
corresponds e.g., to a random field in which the groups $\{{\cal
G}_k\}_{k=1}^{K}$ correspond to different regions affected by
groups of uncorrelated sources. Each of the sub-vectors
$\bbs_{{\cal G}_k}$ in $\bbs_{P}$ contains sources affecting a
certain region in the field and are uncorrelated from the other
sources  present in $\bbs_P$. \textcolor{black}{It is worth
stressing that (a2) does not prevent applicability of the ECD-SPCA
(or BCD-SPCA) algorithm, but it is introduced here only to assess
its asymptotic performance.} Before stating the result proved in
Appendix C, let $\bbv({\cal F})$ denote the entries of vector
$\bbv$ with indices belonging to the set ${\cal F}$.
\textcolor{black}{Further, let ${\cal S}(\bbv)$ denote the support
of $\bbv$, i.e., the set of indices of the nonzero entries of
$\bbv$.}
\begin{proposition}\label{Prop:Ensemble_Noisy}
Let $\bbSigma_x=\bbSigma_s+\bbSigma_{w}$, with $\bbSigma_{s}$
satisfying (a2). Further, assume that the spectral radius of
$\bbSigma_{w}$, namely $d_{\textrm{max}}(\bbSigma_{w})$, satisfies
$d_{\textrm{max}}(\bbSigma_{w})<\Delta(\bbSigma_{s})$, where
$\Delta(\bbSigma_{s})>0$  is a function of $\bbSigma_{s}$. If
$\{\lambda_{\rho}=\lambda\}_{\rho=1}^{q}$ are selected such that $\|{\bbC}_{e,\rho:}\|_0\geq 2$, then
for any arbitrarily small $\delta>0$ there exists a $\mu_o$ such
that for any $\mu\geq \mu_o$ the minimization in
\eqref{Eq:Ens_L1_PCA} admits an optimal solution satisfying
\begin{align}\label{Eq:Support_of_BC}
&\|\bbC_{e,\rho:}^T(\bar{\cal S}_{i_{\rho}})\|_1\leq
\delta,\;\textrm{\emph{and   }} \|\bbC_{e,\rho:}^T({\cal
S}_{i_{\rho}})\|_1\geq \xi(\lambda_{\rho})>0\\
&\|\bbB_{e,:\rho}(\bar{\cal S}_{i_{\rho}})\|_1\leq
\delta,\;\textrm{\emph{and   }} \|\bbB_{e,:\rho}({\cal
S}_{i_{\rho}})\|_1\geq \xi(\lambda_{\rho})>0,\;\;\textrm{ for
}\rho=1,\ldots,q
\end{align}
where $\bar{{\cal S}}_{i_{\rho}}$ is the complement of the support
${\cal S}_{i_{\rho}}$ of $\bbu_{s,i_{\rho}}$,
 while $\{i_{1},\ldots,i_{q}\}\subseteq\{1,\ldots,r\}$. The
 constant $\xi_{\rho}(\lambda_{\rho})$ depends only on
 $\lambda_{\rho}$ and is strictly positive for a finite
 $\lambda_{\rho}$.
\end{proposition}

Prop. \ref{Prop:Ensemble_Noisy} asserts that for $n$ sufficiently
large, \textcolor{black}{S-PCA has an optimal solution $({\bbB}_{e}, {\bbC}_{e})$ whose support is a subset of the true support of
$\{\bbu_{s,i_{\rho=1}}\}_{i=1}^{q}$ even in the presence of
colored noise}. This is possible since for the $\rho$th row of
$\bbC_{e}$, there is a corresponding $i_{\rho}$ column of matrix
$\bbU_{s,r}$ such that $\|\bbC_{e,\rho:}^T(\bar{\cal
S}_{i_{\rho}})\|_1\leq \delta$ for arbitrarily small $\delta$,
while $\|\bbC_{e,\rho:}^T({\cal
S}_{i_{\rho}})\|_1\geq\xi_{\rho}(\lambda_{\rho})> \delta\geq 0$
(strictly positive). Thus, all the nonzero entries of
$\bbC_{e,\rho:}^T$ with magnitude exceeding $\delta$ will have
indices in ${\cal
S}_{i_{\rho}}:=\textrm{support}(\bbu_{s,i_{\rho}})$.  This happens
since: i) $\|\bbC_{e,\rho:}^T(\bar{\cal S}_{i_{\rho}})\|_1$ can be
made arbitrarily small, thus all entries of $\bbC_{e,\rho:}^T$
with indices in $\bar{\cal S}_{i_{\rho}}$ can be driven
arbitrarily close ($\delta$-close) to zero by controlling $\mu$;
and ii) $\|\bbC_{e,\rho:}^T({\cal S}_{i_{\rho}})\|_1$ is strictly
positive with $\xi_{\rho}(\lambda_{\rho})>\delta$, thus some of
the entries of $\bbC_{e,\rho:}^T$ with indices in ${\cal
S}_{i_{\rho}}$ must have magnitude greater than $\delta$. The
number of nonzero entries in $\bbC_{e,\rho:}^T({\cal
S}_{i_{\rho}})$ is determined by $\lambda_{\rho}$. Thus, if
$\lambda_{\rho}$ is selected such that
$\|\bbC_{e,\rho:}^T\|_0=\|\bbu_{s,i_{\rho}}\|_{0}$, then recovery
of the whole support ${\cal S}_{i_{\rho}}$ is ensured.

\textbf{Remark 2:} It should be clarified that the vectors
$\{\bbu_{s,i_{\rho}}\}_{\rho=1}^{q}$ in Prop.
\ref{Prop:Ensemble_Noisy} may not all correspond to the $q$
principal eigenvectors of $\bbSigma_{s}$. Nonetheless S-PCA has an
edge over standard PCA  when colored noise corrupts the training
data. If the observation noise is white, the eigenspaces of
$\bbSigma_x$ and $\bbSigma_s$ coincide and the standard PCA  will
return the $q$ principal eigenvectors of $\bbSigma_s$. However, if
$\bbw_t$ is colored the $q$ principal eigenvectors of
$\bbSigma_x$, namely $\{\bbu_{x,\rho}\}_{\rho=1}^{q}$, will be
different from $\{\bbu_{s,\rho}\}_{\rho=1}^{q}$ and may not be
sparse. Actually, in \emph{standard PCA} (cf. $\lambda_{\rho}=0$)
the magnitude of $\|\bbC_{e,\rho:}^T(\bar{\cal
S}_{i_{\rho}})\|_{1}$ depends on  $\bbSigma_w$ and cannot be made
arbitrarily small. Thus, the magnitude between the entries of
$\bbC_{e,\rho:}^T$  with indices in $\bar{\cal S}_{i_{\rho}}$
relative to those those with indices in ${\cal S}_{i_{\rho}}$
cannot be controlled, for a given noise covariance matrix. This
prevents one from discerning zero from nonzero entries in
$\bbC_{e,\rho:}^T$, meaning that standard PCA cannot guarantee
recovery even of a subset of the support of $\bbu_{s,i_{\rho}} $.

On the other hand, Prop. \ref{Prop:Ensemble_Noisy} states that
S-PCA is capable of identifying a subset of (or all) the support
index set of $\{\bbu_{s,i_{\rho}}\}_{\rho=1}^{q}$.
\textcolor{black}{S-PCA is more resilient to colored noise than
standard PCA because it exploits the sparsity present in the
eigenvectors of $\bbSigma_s$. Intuitively, the $\ell_1$
regularization terms act as prior information facilitating
emergence of the $\bbU_{s,r}$ zero entries in $\bbC_e$ and
$\bbB_e$ as long as the noise variance is not high. Although
$\Delta(\bbSigma_s)$ has not been explicitly quantified, the
upshot of Prop. \ref{Prop:Ensemble_Noisy} is that S-PCA is
expected to estimate better the columns of $\bbU_{s,r}$ when
compared to standard PCA under comparable noise power. Numerical
tests will demonstrate that S-PCA achieves a smaller
reconstruction MSE
even in the presence of colored noise.} 

\subsection{Oracle Properties}\label{Sec:Perf_Anal}

Turning now attention to the noiseless scenario
($\bbSigma_x=\bbSigma_s$), S-PCA is expected to perform
satisfactorily as long as it estimates well the $q$ principal
eigenvectors of $\bbSigma_s$. Reliable estimators of the
clairvoyant matrices $\bbB_o=\bbC_o^T=\bbU_{s,q}^T$ can be
obtained when a growing number of training vectors ensures that:
i) the probability of identifying the zero entries of the
eigenvectors approaches one; and also ii) the estimators of the
non-zero entries of $\bbU_{s,q}$ satisfy a weak form of
consistency \cite{Zou_Ada_Lasso}. Scaling rules for the
$\lambda_{\rho}$'s will be derived to ensure that the S-PCA
estimates $\hat{\bbB}$ and $\hat{\bbC}$ satisfy these so-termed
oracle properties. The forthcoming results will be established for
the BCD-SPCA scheme (of Sec. \ref{Sec:SPCA_BCD}), but similar
arguments can be used to prove related
claims for ECD-SPCA. 

To this end, consider a weighted $\ell_1$-norm in
\eqref{Eq:L1_PCA}, where the sparsity-controlling coefficient
multiplying  $|\bbB(j,\rho)|$ and $|\bbC(\rho,j)|$, namely
$\lambda_{\rho,n}$, is replaced by the product
$\hat{w}_{j,\rho,n}\lambda_{\rho,n}$. Note the dependence of
$\lambda_{\rho,n}$ on $n$, while the $w$'s are set equal to
$\hat{w}_{j,\rho,n}:=|\hat{\bbU}_{s,n}(j,\rho)|^{-\gamma}$,
\textcolor{black}{with $\gamma>0$ and $\hat{\bbU}_s$ denoting the
estimate of $\bbU_s$ obtained via standard PCA.} If
$\bbU_{s}(j,\rho)$ is zero, then for $n$ sufficiently large the
estimate $\hat{\bbU}_{s,n}(j,\rho)$ will have a small magnitude.
This means large weight $\lambda_{\rho,n}\hat{w}_{j,\rho,n}$ and
thus strongly encouraged sparsity in the corresponding estimates
$\hat{\bbB}(j,\rho)$ and $\hat{\bbC}(\rho,j)$. The oracle
properties for $\hat{\bbB}_{\tau,n}$ and $\hat{\bbC}_{\tau,n}$ are
stated next and proved in Appendix D.
\begin{proposition}\label{Prop:Asymptotic_Normality}
Let $\hat{\bbC}_{\tau,n}$ in \eqref{Eq:Btaup1} be an
asymptotically normal estimator of $\bbC_o=\bbU_{s,q}^T$; that is,
\begin{equation}\label{Eq:C_tau_asym_normal}
\hat{\bbC}_{\tau,n}=\bbU_{s,q}^T+\sqrt{n^{-1}}{\bbE}_{\tau,n}^{c}
\end{equation}
where the $j$th column of ${\bbE}_{\tau,n}^c$ converges in
{distribution}, as $n\rightarrow\infty$, to ${\cal
N}(\mathbf{0},\bbSigma_{E_c,j})$, i.e., a zero-mean Gaussian with
covariance $\bbSigma_{E_c,j}$. If the sparsity-controlling
coefficients are chosen so that
\begin{equation}\label{Eq:Scaling_Law_lambda}
\lim_{n\rightarrow\infty}
\frac{\lambda_{\rho,n}}{\sqrt{n}}=0,\;\;\; \textrm{ and
}\lim_{n\rightarrow\infty}\frac{\lambda_{\rho,n}}{n^{(1-\gamma)/2}}=\infty
\end{equation}
then it holds under (a1) that \eqref{Eq:Btaup1} yields an
{asymptotically normal} estimator of $\bbB_o=\bbU_{s,q}$; that is,
\begin{equation}\label{Eq:B_tau_asym_normal}
\hat{\bbB}_{\tau,n}=\bbU_{s,q}+\sqrt{n^{-1}}{\bbE}_{\tau,n}^{b}
\end{equation}
where $\textrm{vec}({\bbE}_{\tau,n}^b)({\cal
S}_o)\xrightarrow[n\rightarrow\infty]{d}{\cal
N}(\mathbf{0},[\bbSigma_{E_b}]_{{\cal S}_{o}})$ (convergence in
distribution); $[\bbSigma_{E_b}]_{{\cal S}_o}$ denotes the
submatrix formed by the rows and columns, with indices in ${\cal
S}_o:={\cal S}(\bbU_{s,q})={\cal
S}(\textrm{\emph{vec}}(\bbU_{s,q}))$, of the error covariance
$\bbSigma_{E_b}$ of $\textrm{vec}({\bbE}_{\tau,n}^b)$
obtained when $\hat{\bbB}_{\tau,n}$ is evaluated by standard PCA.
It follows that
\begin{equation}\label{Eq:Prob_support}
\lim_{n\rightarrow\infty}\textrm{\emph{Pr}}[{\cal
S}(\hat{\bbB}_{\tau,n})={\cal S}(\bbU_{s,q})]=1.
\end{equation}
\end{proposition}

\noindent Following similar arguments as in Prop.
\ref{Prop:Asymptotic_Normality}, it is possible to establish the
following corollary.
\begin{corollary}\label{Cor:Asymptotic_Normality}
If $\hat{\bbB}_{\tau,n}$ is an asymptotically normal estimator of
$\bbB_o$, and the sparsity-controlling coefficients are selected
as in Prop. \ref{Prop:Asymptotic_Normality}, then the optimal
solution of \eqref{Eq:C_tauplus1} at iteration $\tau+1$, namely
$\hat{\bbC}_{\tau+1,n}$, is an asymptotically normal estimator of
$\bbC_o$ and $\lim_{n\rightarrow\infty}\textrm{\emph{Pr}}[{\cal
S}(\hat{\bbC}_{\tau+1,n})={\cal S}(\bbU_{s,q}^T)]=1$.
\end{corollary}
Prop. \ref{Prop:Asymptotic_Normality} and Corollary
\ref{Cor:Asymptotic_Normality} show that when the BCD-SPCA (or
ECD-SPCA) is initialized properly and the sparsity-controlling
coefficients follow the scaling rule in
\eqref{Eq:Scaling_Law_lambda}, then the iterates
$\hat{\bbB}_{\tau,n}$ and $\hat{\bbC}_{\tau,n}$ satisfy the oracle
properties for any iteration index $\tau$. This is important since
it shows that the sparsity-aware estimators $\hat{\bbB}_{\tau,n}$
and $\hat{\bbC}_{\tau,n}$ achieve  MSE performance which
asymptotically is as accurate as that attained by a standard PCA
approach for the nonzero entries of $\bbU_{s,q}$. This holds since
the error covariance matrix of the estimates for the nonzero
entries of $\bbU_{s,q}$, namely $[\bbSigma_{E_b}]_{{\cal S}_{o}}$,
coincides with that corresponding to the standard PCA. The
estimator $\hat{\bbB}_{0,n}=\hat{\bbC}_{0,n}^T=\hat{\bbU}_{s,q}$
obtained via standard PCA and used to initialize the BCD-SPCA and
ECD-SPCA, is asymptotically normal \cite{Brillinger_Time_Series,
Jolliffe_PCA_Book}.

\textbf{Remark 3:} The scaling laws in
\eqref{Eq:Scaling_Law_lambda} resemble those  in \cite[Thm.
2]{Zou_Ada_Lasso} for a linear regression problem. The difference
here is that the estimate $\hat{\bbB}_{\tau-1,n}$ (or
$\hat{\bbC}_{\tau,n}$) is nonlinearly related with
$\hat{\bbC}_{\tau,n}$ ($\hat{\bbB}_{\tau,n}$ respectively). Thus,
establishing Prop. \ref{Prop:Asymptotic_Normality} requires extra
steps to account for the nonlinear interaction between
$\hat{\bbC}_{\tau,n}$ and $\hat{\bbB}_{\tau,n}$. \textcolor{black}{
In order to show asymptotic normality, the chosen weights
$\hat{w}_{j,\rho,n}$ are not that crucial. Actually, the part of
the proof in Apdx. D that establishes asymptotic normality is
valid also when, e.g., $\hat{w}_{j,\rho,n}=1$ and
$\lim_{n\rightarrow\infty}{{n}}^{-1/2}{{\lambda_{\rho,n}}}=0$.
However, the proposed weights $\hat{w}_{j,\rho,n}$ are
instrumental when proving that the probability of recovering the
ground-truth support of $\bbU_{s,q}$ converges to one as the
number of training data grows large.}

Although the probability of finding  the correct support goes to
one asymptotically as $n\rightarrow\infty$, numerical tests
indicate that this probability is high even when $n$ and $\tau$
are finite. This is not the case for the standard PCA estimator,
namely $\hat{\bbU}_{s,q,n}=\textrm{{q-principal
eigenvecs}}(n^{-1}\sum_{t=1}^n\bbs_t\bbs_t^T)$.
\textcolor{black}{Numerical examples will also demonstrate that
even for a finite number  of training data $n$, the probability of
identifying the correct support is increasing as the coordinate
descent iteration index $\tau$ increases.} Thus, the signal
subspace estimates obtained by BCD-SPCA (as well as ECD-SPCA) are
capable of yielding the correct support of $\bbU_{s,q}$ even when
$n$ is sufficiently large but finite. Consequently, improved
estimates of $\bbU_{s,q}$ are obtained which explains the lower
$J_{\textrm{rec}}(\bbB,\bbC)$ attained by S-PCA relative to PCA.



\section{S-PCA Based transform coding}\label{Sec:STC}

Up to this point sparsity has been exploited for DR of data
vectors with analog-amplitude entries. However, the majority of
modern compression systems are digital. This motivates
incorporation of sparsity also in the quantization module that
follows DR. This two-stage process comprises the transform coding
(TC) approach which has been heavily employed in image compression
applications due to its affordable computational complexity
\cite{Transform_Coding}. However, current TC schemes do not
exploit the presence of sparsity that may be present in the
covariance domain.

A sparsity-aware TC (SATC) is proposed here to complement the
BCD-SPCA (or ECD-SPCA) algorithm during the data transformation
step. The basic idea is to simply quantize the DR vectors using a
VQ. Given $\bbx$, the DR matrix $\hat{\bbC}$ obtained by S-PCA is
employed during the transformation step to produce the DR vector
$\bby=\hat{\bbC}\bbx$. Then, VQ is employed to produce at the
output of the encoder a vector of quantized entries
$\hat{\bbeta}_{Q}=Q[\bby]\in{\cal C}_{Q}^{q\times 1}$, where
${\cal C}_{Q}:=\{\hat{\bbeta}_1,\ldots,\hat{\bbeta}_L\}$ is the
quantizer codebook with cardinality $L=2^R$, where $R$ denotes the
number of bits used to quantize $\bby$. The VQ will be designed
numerically using the Max-Lloyd algorithm, as detailed in e.g.
\cite{Quantization_Book}, which uses
$\{\bby_{t}=\hat{\bbC}\bbx_{t}\}_{t=1}^{n}$ to determine the
quantization cells $\{{\cal R}_{\hat{\bbeta}_{l}}\}_{l=1}^{L}$,
and their corresponding centroids, a.k.a. codewords
$\{\hat{\bbeta_{l}}\}_{l=1}^{L}$.

During decoding, the standard process in typical TC schemes
\cite{Quantization_Book,Transform_Coding} is to multiply
$\hat{\bbeta}_Q=Q[\hat{\bbC}\bbx]$ with the matrix
$(\hat{\bbC}^T\hat{\bbC})^{\dag}\hat{\bbC}^T$, and form the estimate
$\hat{\bbs}=(\hat{\bbC}^T\hat{\bbC})^{\dag}\hat{\bbC}^T\hat{\bbeta}_{Q}$.
This estimate minimizes the Euclidean distance
$\|\hat{\bbeta}_{Q}-\hat{\bbC}\bbu\|$ wrt $\bbu$. Note that the
reconstruction stage of SATC is also used in the DR setting
considered in Sections \ref{Sec:Prelims_Form} and \ref{Sec:SPCA},
except that $\hat{\bbeta}_{Q}$ is replaced with the vector
$\hat{\bbC}\bbx$ whose entries are analog. \textcolor{black}{The
reason behind using only $\hat{\bbC}$, and not $\hat{\bbB}$, is the
penalty term $\|\bbB-\bbC^T\|_{F}^2$ which ensures that $\hat{\bbC}$
and $\hat{\bbB}^T$ will be close in the $\ell_2$-error sense.
Certainly, $\hat{\bbB}$ could have been used instead, but such a
change would not alter noticeably the reconstruction performance.}
Simulations will demonstrate that the sparsity-inducing mechanisms
in the DR step assist SATC to achieve improved MSE reconstruction
performance when compared to related sparsity-agnostic TCs.

\section{Simulated Tests}\label{Sec:Sim}
Here the reconstruction performance of ECD-SPCA is studied and
compared with the one achieved by standard PCA, as well as
sparsity-aware alternatives that were modified to fit the
dimensionality reduction setting. The different approaches are
compared both in the noiseless and noisy scenarios. Simulation tests
are also performed to corroborate the oracle properties established
in Sec. \ref{Sec:Perf_Anal}. The SATC is compared with conventional
TCs in terms of reconstruction MSE using synthetic data first. Then,
SATC is tested in an image compression and denoising application
using images from \cite{Nasa_Martian_Images}.

\subsection{Synthetic Examples}\label{Sec:Synthetic_Sim}

\begin{figure}
\begin{center}
\centerline{\epsfig{file=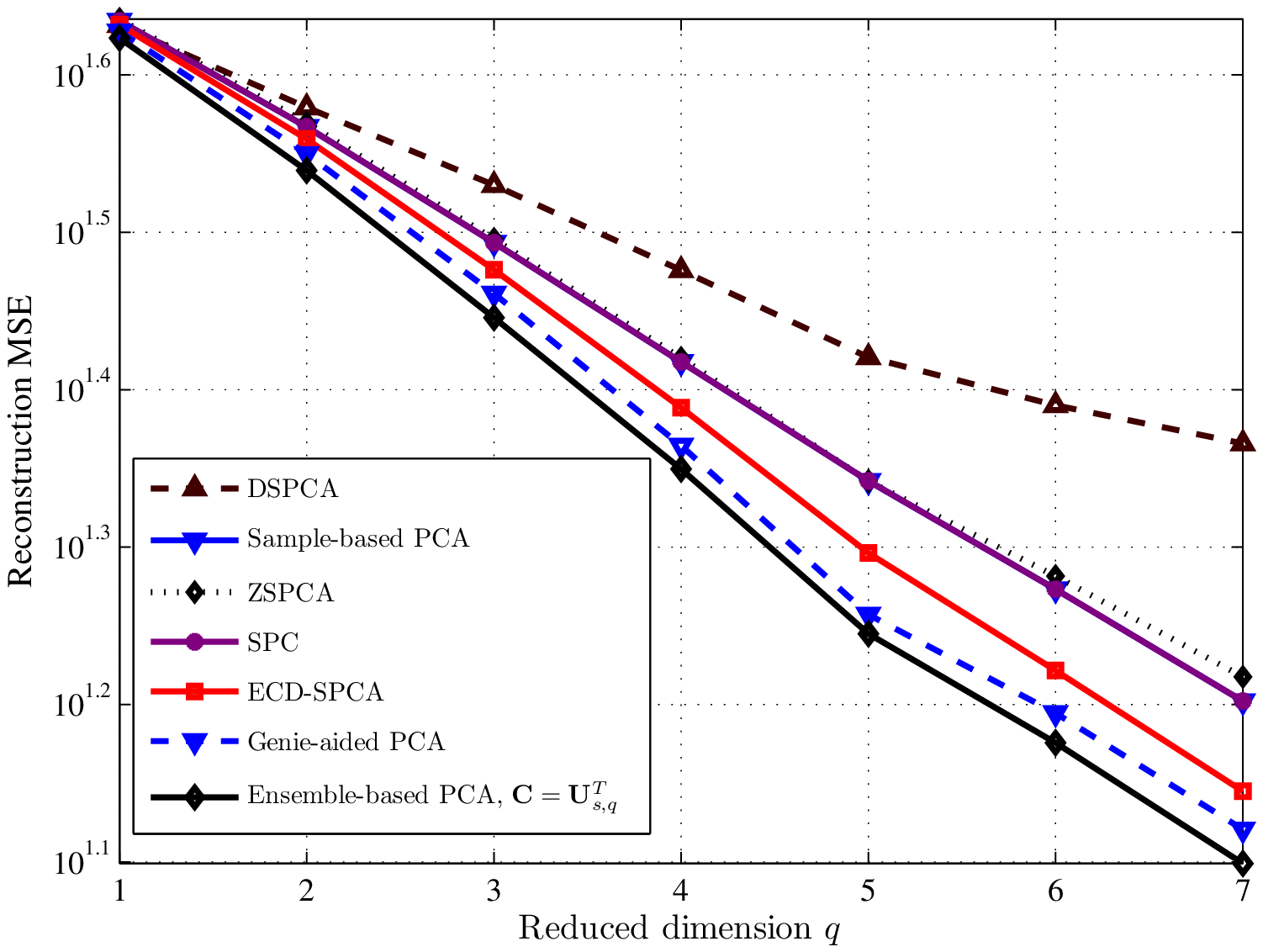,width=3.9in}\hspace*{-0.2cm}
\epsfig{file=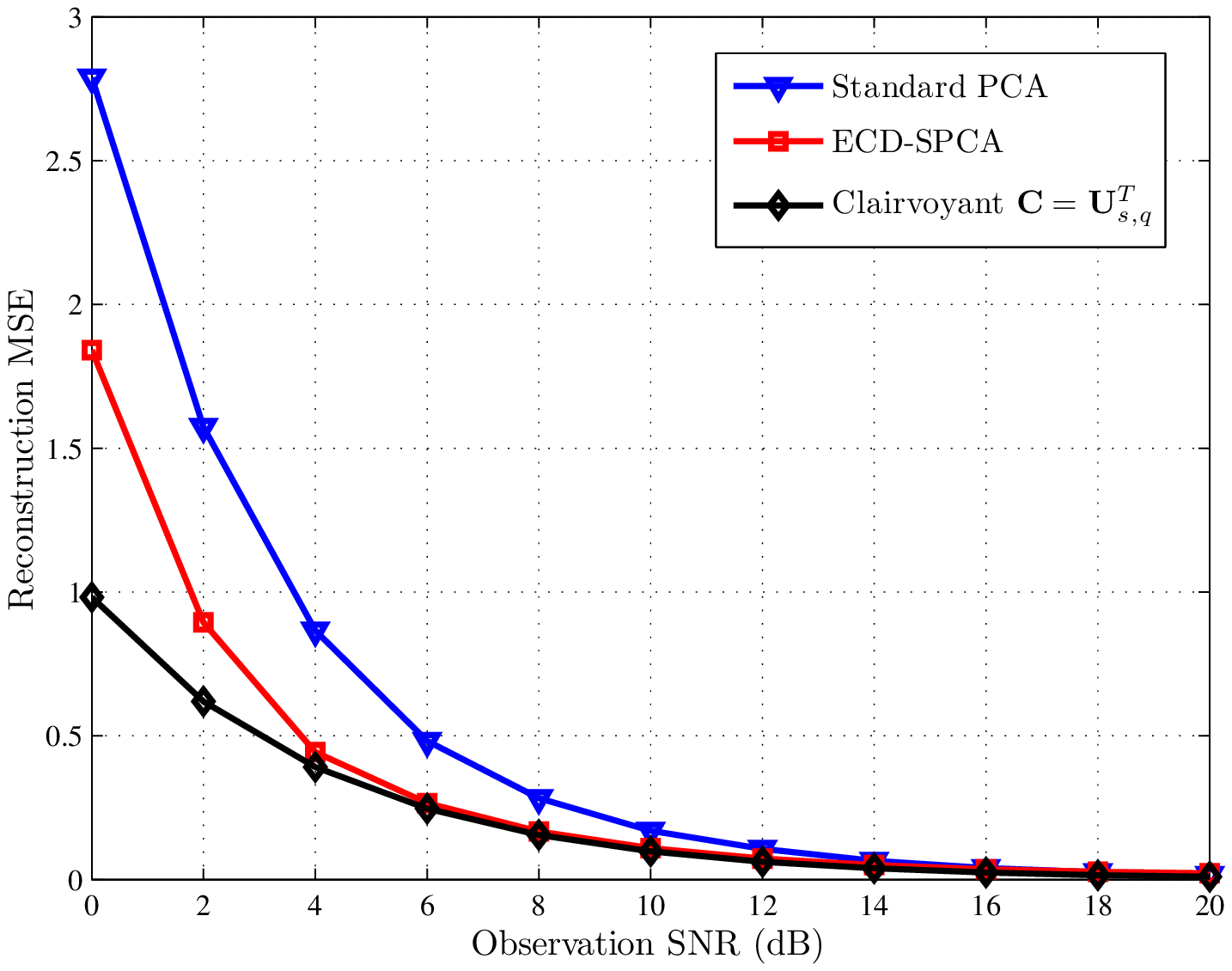,width=3.9in}}\vspace{-0.5cm}
\caption{Reconstruction MSE vs. $q$ in the noiseless case (left);
and observation SNR in the noisy scenario with $q=3$ (right).}
\label{Fig:Rec_Mse_vs_q}
\end{center}
\end{figure}

The reconstruction MSE $J_{\textrm{rec}}(\bbB_m,\bbC_m)$ is
measured for matrices ${\bbB}_m$ and ${\bbC}_m$ obtained via: i)
ECD-SPCA; ii) the true signal subspace, i.e.,
$\mathbf{B}_o=\mathbf{C}_o^T=\bbU_{s,q}$; iii) a `sample'-based
PCA approach where $\hat{\bbU}_{x,q}$ is used; iv) a genie-aided
PCA which relies on (iii) but also knows where the zero entries of
$\mathbf{U}_{s,q}$ are located; v) the sparse PCA approach in
\cite{Zhou_Spca} abbreviated as ZSPCA; vi) the scheme in
\cite{Witten_Spc} abbreviated as SPC; and vii) the algorithm of
\cite{Aspremont_Dspca}, which is abbreviated as DSPCA. With $p=14$
and $n=50$, the MSEs throughout the section are averaged over
$200$ Monte Carlo runs using a data set that is different from the
training set $\bbX$. In the noiseless case,
$\bbSigma_x=\bbSigma_s$ is constructed \textcolor{black}{to be a
permuted block diagonal matrix} with $r=8$, while $80\%$ of the
entries of $\bbU_{s,r}$ are zero. \textcolor{black}{The
sparsity-controlling coefficients multiplying $|\bbB(j,\rho)|$ and
$|\bbC(\rho,j)|$ are set equal to
${\lambda_{\rho,n}}|\hat{\bbU}_{x}(i,j)|^{-\gamma}$,  with
$\gamma=1$ and $\lambda_{\rho,n}\sim n^{0.3}$.} Fig.
\ref{Fig:Rec_Mse_vs_q} (left) depicts $J_{\textrm{rec}}(\cdot)$
versus $q$. The sparsity coefficients in the  sparsity-aware
approaches are selected from a search grid to achieve the smallest
possible reconstruction MSE. Clearly ECD-SPCA exploits the
sparsity present in $\bbSigma_s$ and achieves a smaller
reconstruction MSE that is very close to the genie-aided approach.
\textcolor{black}{Note that in Fig. \ref{Fig:Rec_Mse_vs_q} (left)
there are seven curves. The curve corresponding to sample-based
PCA almost overlaps with the one corresponding to SPC. It is also
observed that the more sparse the eigenvectors in $\bbU_{s,r}$
are, the more orthogonal are the ECD-SPCA estimates $\hat{\bbB}$
and $\hat{\bbC}$. This suggests that the $\ell_1$ regularization
terms in S-PCA induce approximate orthogonality in the
corresponding estimates, as long as the underlying eigenvectors
forming $\bbU_{s,r}$ are sufficiently sparse.}

Fig. \ref{Fig:Rec_Mse_vs_q} (right) the reconstruction MSE is
plotted as a function of the observation SNR, namely
$\textrm{SNR}_{\textrm{obs}}:=10\log_{10}[\textrm{tr}
(\bbSigma_s)/\textrm{tr}(\bbSigma_w)]$. The colored noise
covariance matrix is factored as $\bbSigma_w=\bbM_w\bbM_w^T$,
where $\bbM_w$ is randomly generated matrix with Gaussian i.i.d.
entries. The ECD-SPCA scheme is compared with the
sparsity-agnostic standard PCA approach. With $r=q=3$ and $p=14$,
$\bbSigma_s$ is constructed \textcolor{black}{to be a permuted
block diagonal matrix} such that $70\%$ of the entries of the
eigenmatrix $\bbU_{s,r}$ are equal to zero. All sparsity
coefficients in ECD-SPCA are set equal to $\lambda=5*10^{-3}$.
Fig. \ref{Fig:Rec_Mse_vs_q} (right) corroborates that the novel
S-PCA  can lead to better reconstruction/denoising performance
than the standard PCA. The MSE gains are noticeable in the
low-to-medium SNR regime. The sparsity imposing mechanisms of
ECD-SPCA lead to improved subspace estimates yielding a
reconstruction MSE that is close to the one obtained using
$\bbU_{s,q}$. This result corroborates the claims of Prop.
\ref{Prop:Ensemble_Noisy}.

\begin{figure}
\begin{center}
\centerline{\hspace{-1cm}\epsfig{file=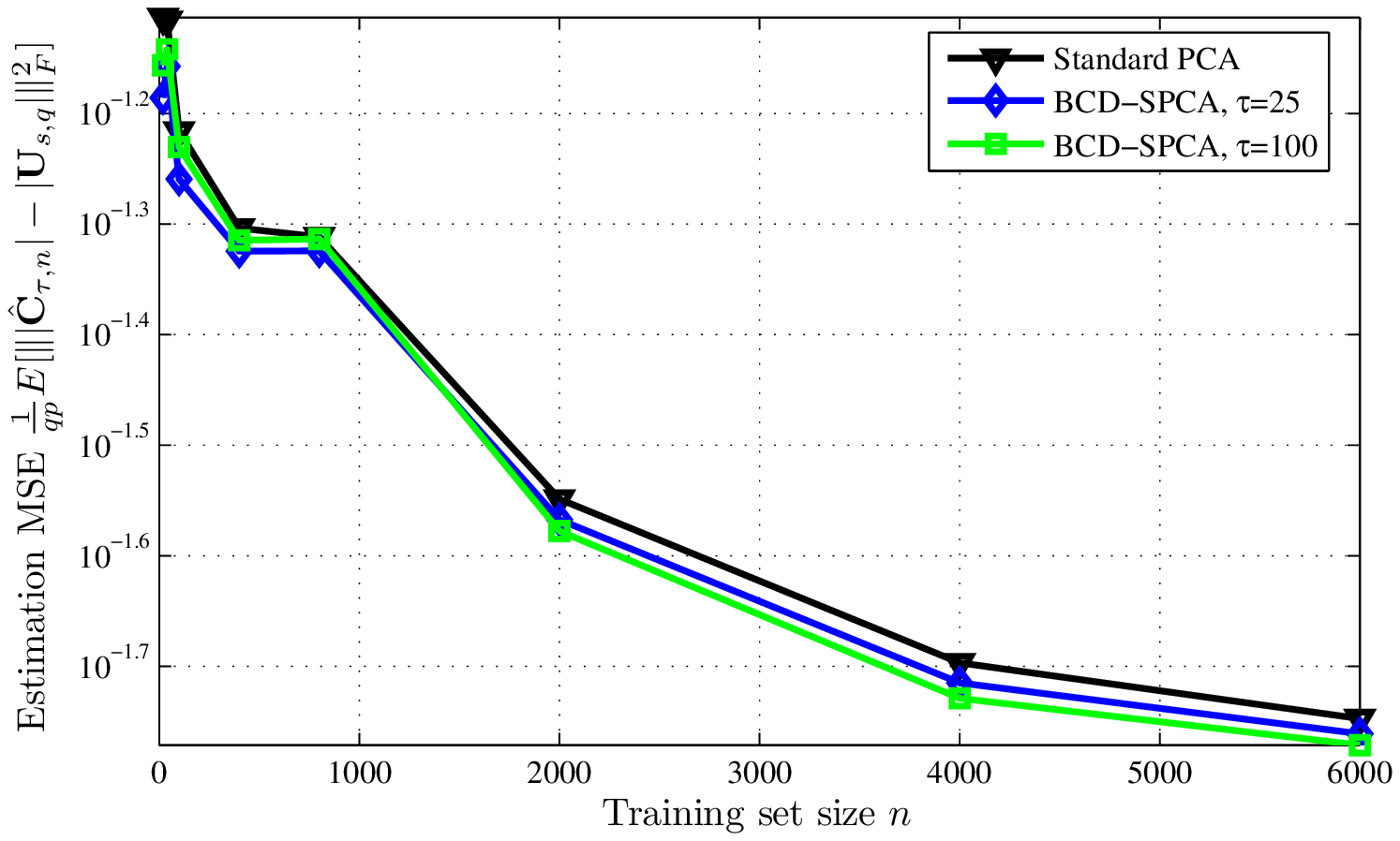,width=4.5in,height=2.7in}
\hspace*{-1.1cm} \epsfig{file=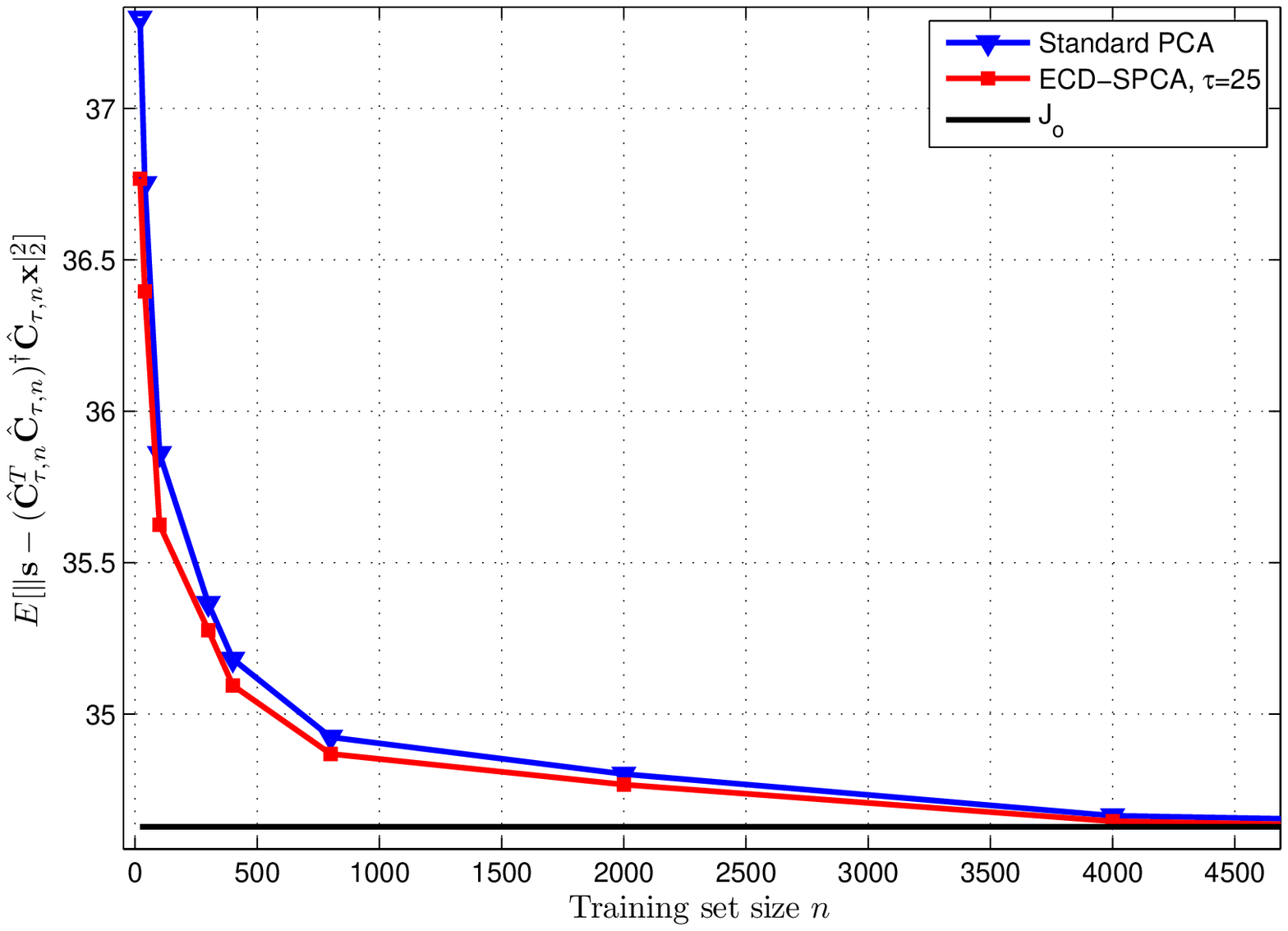,width=4.2in}}
\caption{Normalized estimation MSE
$(qp)^{-1}E[\||\hat{\bbC}_{\tau,n}|-|\bbU_{s,q}|\|_F^2]$ (left);
and reconstruction MSE $J_{\textrm{rec}}$ (right) vs. $n$ for
ECD-SPCA with $p=14$ and $q=2$.} \label{Fig:Est_and_Rec_Mse}
\end{center}
\vspace{-0.6cm}
\end{figure}

\begin{figure}
\centerline{\hspace{0.6cm}\epsfig{file=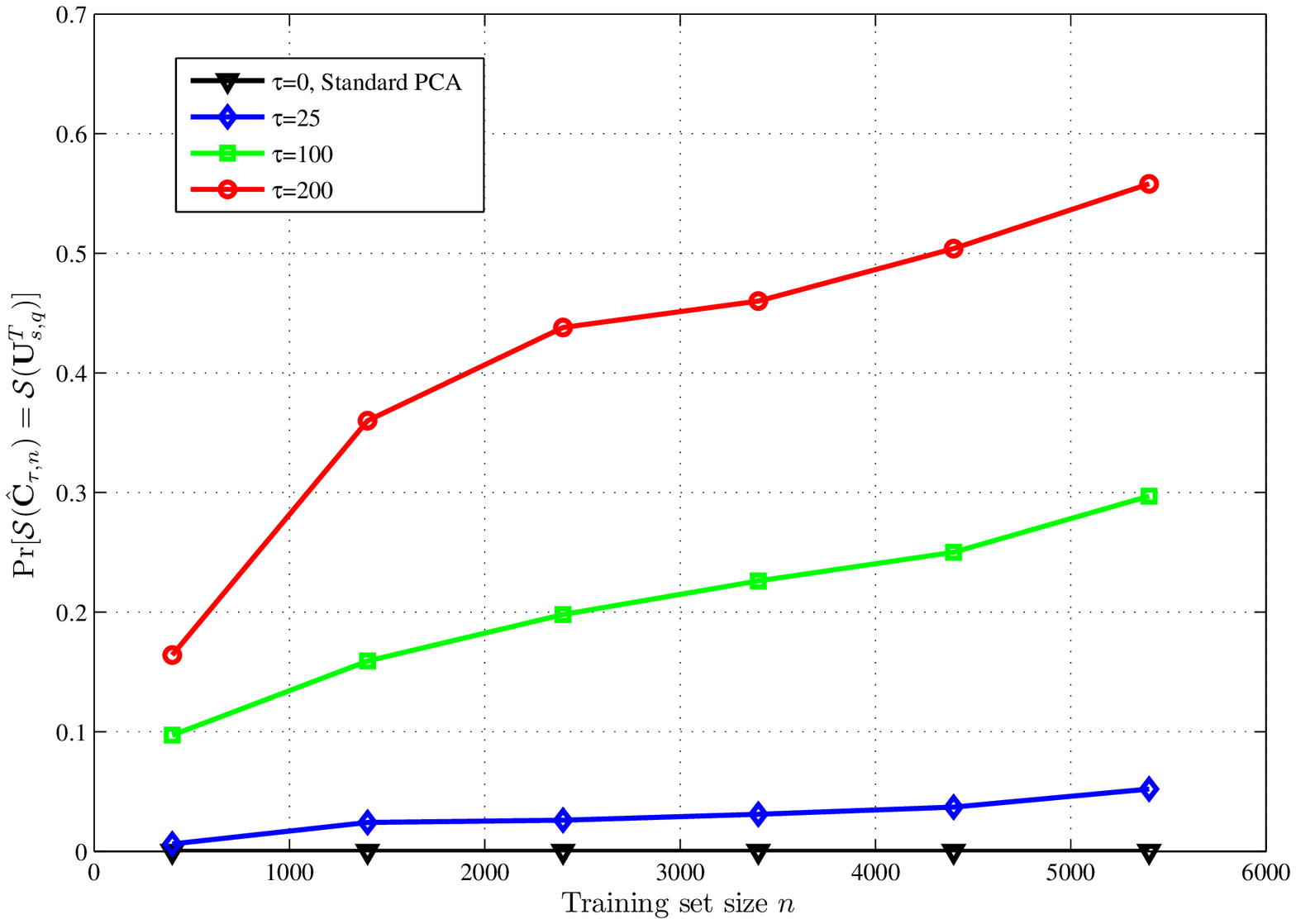,width=4.1in}\hspace{-0.2in}
\epsfig{file=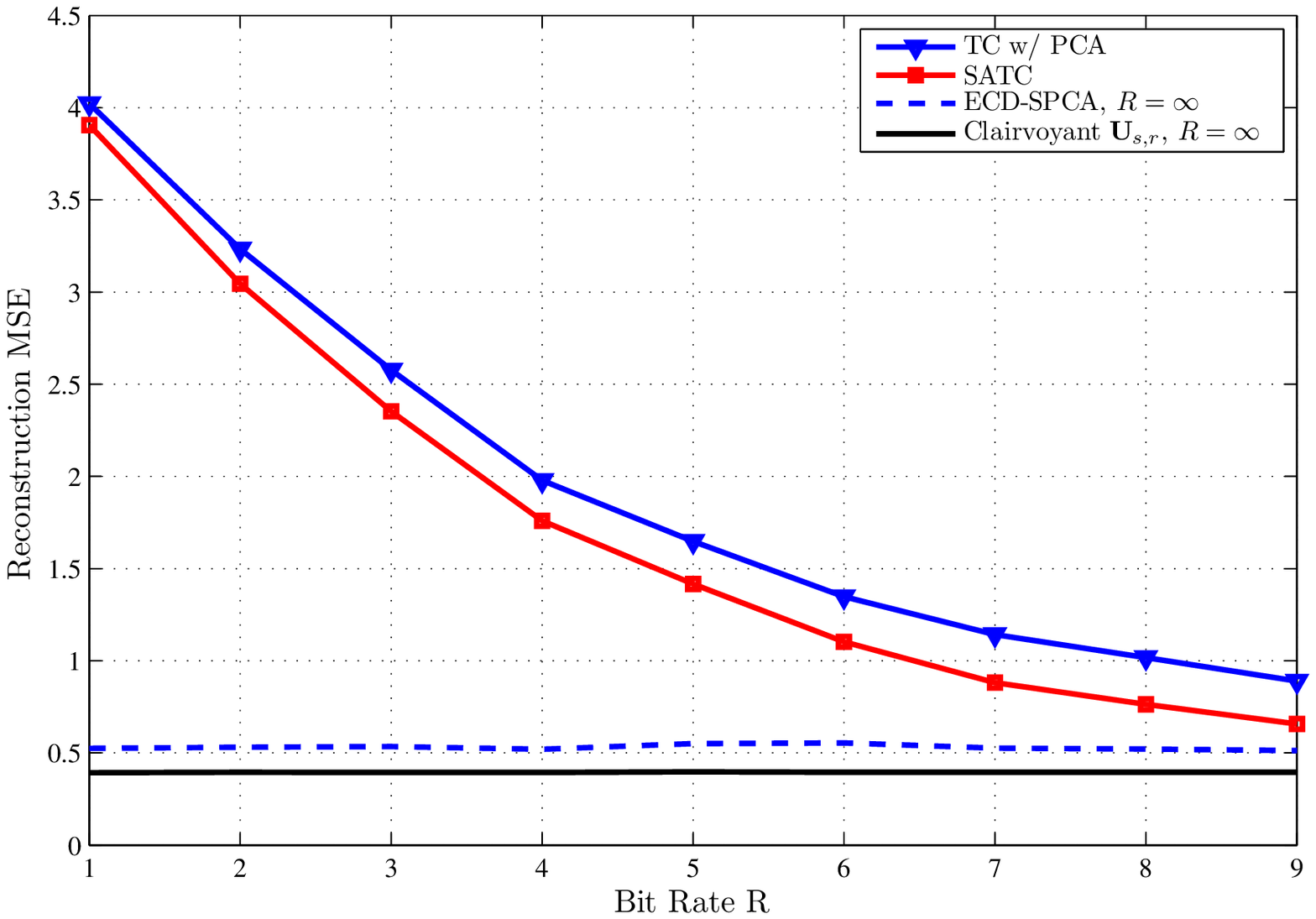,width=4.1in,height=2.82in}
}\vspace{-0.8cm}\caption{(Left) Probability that
$\hat{\bbC}_{\tau,n}$ identifies the support of $\bbU_{s,q}^T$.
Variable-size training  sets considered in the x-axis, while the
different curves correspond to a variable number of coordinate
descent recursions; (Right) Reconstruction MSE vs. quantization
bit rate $R$ for SATC and a PCA-based TC scheme with $r=q=3$.}
\label{Fig:Prob_Support}
\end{figure}

The next three figures validate the S-PCA properties in the
noiseless case (see Sec. \ref{Sec:Perf_Anal}). Consider a setting
where $p=14$, $r=8$, and $q=2$, and $\bbSigma_{x}=\bbSigma_s$
constructed \textcolor{black}{to be a permuted block diagonal matrix}
such that $\bbU_{s,r}$ has $70\%$ of its entries equal to zero.
\textcolor{black}{ The $\lambda$'s are selected as in the first
paragraph.} Fig. \ref{Fig:Est_and_Rec_Mse} (left) displays the
signal subspace estimation MSE
$E[\||\hat{\bbC}_{\tau,n}|-|\bbU_{s,q}^T|\|_F^2]$, where the
$|\cdot|$ operator is applied entry-wise and is used to eliminate
any sign ambiguity present in the rows of $\hat{\bbC}_{\tau,n}$. As
the training data size goes to infinity, the estimation error
converges to zero. The convergence speed is similar to the one
achieved by standard PCA. Similar conclusions can be deduced for the
reconstruction MSE shown in Fig. \ref{Fig:Est_and_Rec_Mse} (right).
The reconstruction MSE associated with ECD-SPCA is smaller than the
one corresponding to standard PCA. The MSE advantage is larger for a
small number of training data in which case standard PCA has trouble
locating the zeros of $\bbU_{s,q}$. These examples corroborate the
validity of Prop. \ref{Prop:Asymptotic_Normality}. Interestingly,
multiple coordinate descent iterations $(\tau>0)$ result in smaller
estimation and reconstruction MSEs than the one achieved by standard
PCA ($\tau=0$). The MSE gains are noticeable for a small number of
training samples. Such gains are expected since ECD-SPCA (or
BCD-SPCA) is capable of estimating the true support ${\cal
S}(\bbU_{s,q})$ with a positive probability even for a finite $n$.
As shown in Fig. \ref{Fig:Prob_Support} (left), for $\tau>0$ the
probability of finding the true support converges to one as
$n\rightarrow\infty$ (cf. Prop. \ref{Prop:Asymptotic_Normality}). As
$\tau$ increases, this probability also increases, while standard
PCA ($\tau=0$) never finds the correct support with a finite number
of training data.

Next, the reconstruction MSE of the SATC (Sec.\ref{Sec:STC}) is
considered and compared with the one achieved by a TC scheme based
on standard PCA. The noisy setting used to generate Fig.
\ref{Fig:Rec_Mse_vs_q} (right) is considered here with $p=14$ and
$n=22$. With $r=q=3$, $\bbSigma_{s}$ is constructed
\textcolor{black}{to be a permuted block diagonal matrix} so that
$80\%$ of the entries of $\bbU_{s,r}$ are zero. Data reduction in
SATC is performed via ECD-SPCA, while its sparsity controlling
coefficients are set as described in the first paragraph of this
section. Once the DR matrix $\hat{\bbC}$ is obtained, the DR
training data $\hat{\bbC}\bbX$ are used to design the VQ using
Max-Lloyd's algorithm. Fig. \ref{Fig:Prob_Support} (right) depicts
the reconstruction MSE versus the number of bits used to quantize
a single DR vector. Fig. \ref{Fig:Prob_Support} (right) clearly
shows that SATC benefits from the presence of sparsity in
$\bbSigma_s$ and achieves improved reconstruction performance when
compared to the standard TC scheme that relies on PCA. The dashed
and solid lines correspond to the reconstruction MSE achieved by
ECD-SPCA and $\bbU_{s,r}$ respectively while no quantization step
is present ($R=\infty$).

\subsection{Image compression and denoising}\label{Sec:Image_Comp}

SATC is tested here for compressing and reconstructing images.
These images have size $180\times 256$ and they are extracted, as
described in Sec.  \ref{Sec:Prelims_Form}, from a bigger image of
size $2520\times 2048$ in \cite{Nasa_Martian_Images}. The images
are corrupted with additive zero-mean Gaussian colored noise
whose covariance $\bbSigma_w$ is structured as
$\bbSigma_{w}=\bbM\bbM^T$, where $\bbM$ contains Gaussian i.i.d.
entries. The trace of  $\bbSigma_w$ is scaled to fix the SNR at
$15$dB. Out of a total of $112$ generated images, $30$ are used
for training to determine the DR matrix $\hat{\bbC}$, and design
the VQ. The rest are used as test images to evaluate the
reconstruction performance of the following three schemes: i) the
SATC; ii) a TC scheme that uses DCT; and iii) a TC scheme which
relies on PCA.

The images are split into blocks of dimension $8\times 8$, and
each of the three aforementioned TC schemes is applied to each
block. Here $p=64$ and $\{\bbx_{t_i}\}_{i=1}^{30}$ is a vectorized
representation of an $8\times 8$ sub-block that consists of
certain image pixels, while $t_i\in\{1,\ldots,112\}$ denotes the
image index. During the operational mode, datum $\bbx$ corresponds
to a noisy sub-block occupying the same row and column indices as
the $\bbx_{t_i}$'s but belonging to an image that is not in the
training set. The signal of interest $\bbs$ corresponds to the
underlying noiseless block we wish to recover. Each noisy datum
$\bbx$ is transformed using either i) the SATC transformation
matrix obtained via ECD-SPCA; or ii) the DCT; or iii) the PCA
matrix. When the DCT is applied, then DR is performed by keeping
the $q$ largest in magnitude entries of the transformed vector.
The reduced dimension here is set to $q=14$. The $q\times 1$
vectors are further quantized  by a VQ designed using the
Max-Lloyd algorithm fed with the DR training vectors. At the
decoder, the quantized vectors are used to reconstruct $\bbs$ by:
i) using the scheme of  Sec. \ref{Sec:STC}; ii) multiplying the
quantized data with $\hat{\bbU}_{x,q}$ (PCA); or iii) applying
inverse DCT  to recover the original block. The
sparsity-controlling coefficients in ECD-SPCA are all set equal to
$\lambda=2*10^{-2}$.

\begin{figure}
\centerline{\epsfig{file=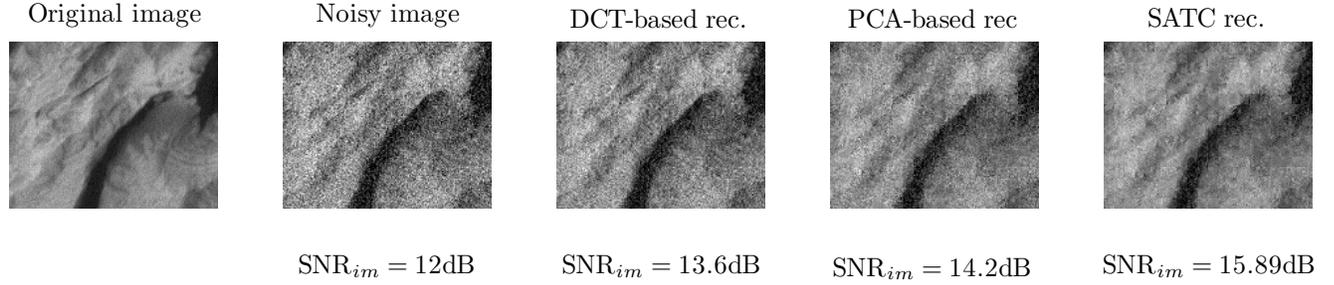}}\vspace{-0.5cm}
\caption{Original image (leftmost); its noisy version; and its
reconstruction using DCT-based TC, PCA-based TC and SATC (right).
Reconstructed images are produced after setting $q=14$ and $R=7$
bits for each of the $8\times 8$ blocks comprising the original
noisy image.} \label{Fig:Martian_Images}
\vspace{-0.6cm}
\end{figure}

\begin{figure}
\begin{center}
\centerline{\epsfig{file=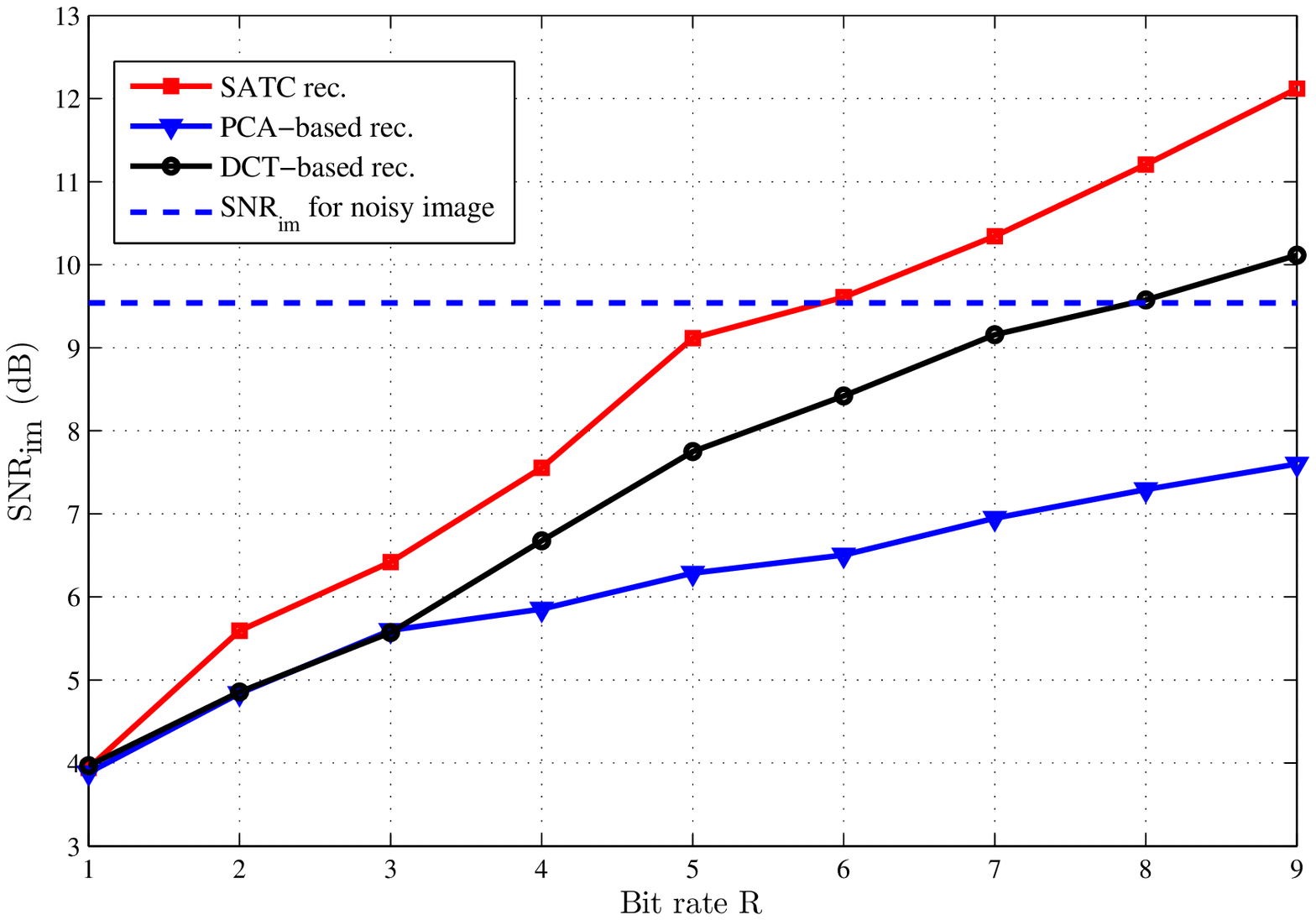,width=3.8in,height=2.69in}\hspace*{-0.2in}
\epsfig{file=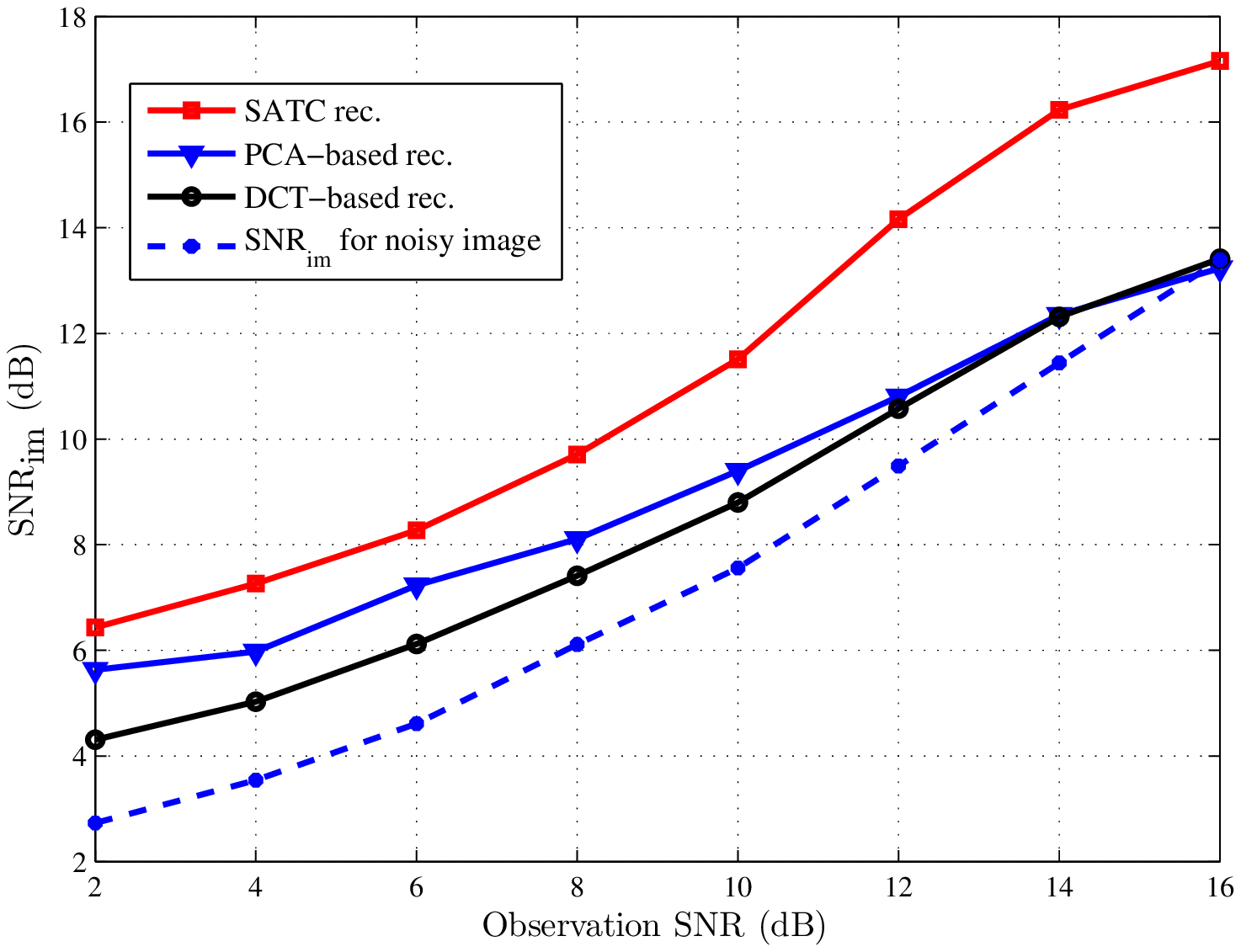,width=3.8in}}\vspace{-0.8cm}
\caption{$\textrm{SNR}_{\textrm{im}}$ vs.  bit rate $R$ (left);
and observation SNR (right) for different TCs using the images
extracted from \cite{Nasa_Martian_Images}.}
\label{Fig:PSNR_vs_bitrate}
\end{center}
\end{figure}

Fig. \ref{Fig:Martian_Images} shows: a) the original image; b) its
noisy version; c) a reconstruction using DCT; d) a reconstruction
using a PCA-based  TC; and e) the reconstruction returned by SATC.
The reconstructed images are obtained after setting the bit rate  to
$R=7$. The reconstruction returned by SATC is visually more pleasing
than the one obtained by the DCT- and PCA-based TCs. The figure of
merit used next is $\textrm{SNR}_{\textrm{im}}:=10log_{10}
[\frac{P_{\textrm{signal}}}{P_{\textrm{error}}}]$, where
$P_{\textrm{signal}}$ denotes the power of the noiseless image and
$P_{\textrm{error}}$ the power of the noise present in the
reconstructed image. Fig. \ref{Fig:PSNR_vs_bitrate} (left) displays
$\textrm{SNR}_{\textrm{im}}$ versus the bit-rate of the VQ with SNR
set at $17$dB. Clearly, SATC achieves higher
$\textrm{SNR}_{\textrm{im}}$ values compared to the DCT- and
PCA-based TCs. SATC performs better because the ECD-SPCA algorithm
used to evaluate the DR matrix $\hat{\bbC}$ takes advantage of the
sparsity present in $\bbSigma_s$. Similar conclusions can be drawn
from Fig. \ref{Fig:PSNR_vs_bitrate} (right) depicting
$\textrm{SNR}_{im}$ versus the  SNR for $q=14$ and $R=7$ bits.

\section{Concluding Remarks}\label{Sec:Conclusion}

The present work dealt with compression, reconstruction, and
denoising of signal vectors spanned by an orthogonal set of sparse
basis vectors that further result a covariance matrix with sparse
eigenvectors. Based on noisy training data,  a sparsity-aware DR
scheme was developed using $\ell_1$-norm regularization to form
improved estimates of the signal subspace leading to improved
reconstruction performance. Efficient coordinate descent
algorithms were developed to minimize the associated non-convex
cost. The proposed schemes were guaranteed to converge at least to
a stationary point of the cost.

Interesting analytical properties were established for the novel
signal subspace estimator showing that even when the noise
covariance matrix is unknown, a sufficiently large signal-to-noise
ratio ensures that the proposed estimators identify (at least a
subset) of the unknown support of the signal covariance
eigenvectors. These results advocate that sparsity-aware compression
performs well especially when a limited number of training data is
available. Asymptotic normality is also established for the
sparsity-aware subspace estimators, while it is shown that the
probability of these estimates identifying the true signal subspace
support approaches one as the number of training data grows large.
Appropriate scaling laws for the sparsity-controlling coefficients
were derived to satisfy the aforementioned properties.

Finally, the novel S-PCA approach was combined with vector
quantization to form a sparsity-aware transform codec (SATC) that
was demonstrated to outperform existing sparsity-agnostic
approaches. Simulations using both synthetic data and images
corroborated the analytical findings and validated the effectiveness
of the proposed schemes. Work is underway to extend the proposed
framework to settings involving compression of nonstationary
signals, and processes with memory.

\section*{Acknowledgment}
The authors would like to thank Prof. N. Sidiropoulos of the
Technical University of Crete, Greece, for his valuable input and
suggestions on the themes of this paper.

{\Large\appendix}

\noindent \emph{A. Proof of Lemma \ref{Lem:Scalar_Lasso}:} The
minimization problem in \eqref{Eq:Scalar_Lasso} can
 be equivalently expressed as
\begin{equation}\label{Eq:Scalar_Lasso_Constrained}
\hat{c}=\arg\min_{c}\|\bbchi-c\bbh\|_2^2+\mu(c-\hat{b})^2+\lambda
t,\;\;\; \textrm{s. to  }-t\leq c\leq t
\end{equation}
and the derivative of its Lagrangian function involving
multipliers $\nu_1$ and $\nu_2$ is given by
$$\nabla_c{\cal L}(c,\nu_1,\nu_2)=
2c\bbh^T\bbh-2c\bbh^T\bbchi+2c\mu-2\mu \hat{b}+\nu_1-\nu_2.$$
After using the KKT necessary optimality conditions \cite[pg.
316]{Bertsekas_Nonlinear_Book}, it can be readily deduced that the
optimal solution of \eqref{Eq:Scalar_Lasso} is given by the second
equation in Lemma \ref{Lem:Scalar_Lasso}.\myQED

\noindent \emph{B. Proof of convergence of ECD-SPCA}

\textcolor{black}{Let
$f(\{\bbB(j,\rho),\bbC(\rho,j)\}_{j=1,\rho=1}^{p,q})$ denote the
S-PCA cost given in \eqref{Eq:L1_PCA}, defined over
$\mathbb{R}^{2pq\times 1}$; and
$f_0(\{\bbB(j,\rho),$
$\bbC(\rho,j)\}_{j=1,\rho=1}^{p,q}):=n^{-1}\|\bbX-\bbB\bbC\bbX\|_F^2+\mu\|\bbB-\bbC^T\|_F^2$.
Next, consider the level set
\begin{equation}\label{Eq:Level_set}
\hspace{-0.45cm}{\cal
F}^0:=\{\{\bbB(j,\rho),\bbC(\rho,j)\}_{j=1,\rho=1}^{p,q}:\;\;
f(\{\bbB(j,\rho),\bbC(\rho,j)\}_{j=1,\rho=1}^{p,q})\leq
f(\hat{\bbB}_{0},\hat{\bbC}_0)\}
\end{equation}
where $\hat{\bbB}_{0}=\hat{\bbC}_0^T=\hat{\bbU}_{s,q}$ correspond
to the matrices used to initialize ECD-SPCA obtained via standard
PCA. If  $\hat{\bbB}_{0}$ and $\hat{\bbC}_{0}$ have  finite
$\ell_1$-norms, the set ${\cal F}^0$ is closed and bounded
(compact). The latter property can be deduced from
\eqref{Eq:L1_PCA} and \eqref{Eq:Level_set}, which ensure that
matrices $\bbB$ and $\bbC$ in ${\cal F}^0$ satisfy
$\sum_{\rho,j}\lambda_{\rho}(|\bbB(j,\rho)|+|\bbC(\rho,j)|\leq
f(\hat{\bbB}_{0},\hat{\bbC}_0)$. Moreover,
$f(\hat{\bbB}_{0},\hat{\bbC}_0)$ is finite when $\hat{\bbB}_{0}$
and $\hat{\bbC}_{0}$ have finite norms. This is true when the
training data $\bbX$ contain finite entries. Thus, ${\cal F}^0$ is
a compact set. Further, the cost function $f(\cdot)$ is continuous
on ${\cal F}^{0}$.}

\textcolor{black} {From \eqref{Eq:Elementwise_minimization} and
\eqref{Eq:Btau_row_l_col_j} it follows readily that the
minimization problems solved to obtain $\hat{\bbC}_{\tau}(j,\rho)$
and $\hat{\bbB}_{\tau}(\rho,j)$, respectively, are strictly
convex. Thus,  minimizing $f(\cdot)$ with respect to an entry of
$\bbC$ or $\bbB$ yields a unique minimizer, namely
$\hat{\bbC}_{\tau}(\rho,j)$, or $\hat{\bbB}_{\tau}(j,\rho)$.
Finally, $f(\cdot)$ satisfies the regularization conditions
outlined in \cite[(A1)]{BCD_Tseng}. Specifically, the domain of
$f_0(\cdot)$ is formed by matrices whose entries satisfy
$\bbB(j,\rho)\in(-\infty,+\infty)$ and
$\bbC(\rho,j)\in(-\infty,+\infty)$ for $j=1,\ldots,p$ and
$\rho=1,\ldots,q$. Thus,
$\textrm{domain}(f_0)=(-\infty,\infty)^{2qp}$ is an \emph{open
set}. Moreover, $f_0(\cdot)$ is G$\hat{\textrm{a}}$uteaux
differentiable over $\textrm{domain}(f_0)$. Specifically, the
G$\hat{\textrm{a}}$uteaux derivative of $f_0(\cdot)$ is defined as
$$f_0'(\bbB,\bbC;\bbDelta_{B},\bbDelta_{C}):=\lim_{h\rightarrow 0}
\frac{f_0(\bbB+h\bbDelta_{B},\bbC+h\bbDelta_{C})-f_0(\bbB,\bbC)}{h}.$$
Applying simple algebraic manipulations it follows readily that
the G$\hat{\textrm{a}}$uteaux derivative exists for all
$\bbDelta_{B},\bbDelta_{C}\in\textrm{domain}(f_0)$, and is equal
to $2\mu\textrm{tr}[(\bbB-\bbC^T)(\bbDelta_B-\bbDelta_C^T)^T]
-2\textrm{tr}[(\bbX-\bbB\bbC\bbX)(\bbDelta_B\bbC\bbX+\bbB\bbDelta_C\bbX)^T]$.
Then, convergence of  the ECD-SPCA iterates to a stationary point
of the S-PCA cost $f(\cdot)$ is readily established using
\cite[Thm. 4.1 (c)]{BCD_Tseng}. \myQED}

\noindent \textcolor{black}{\emph{C. Proof of Proposition
\ref{Prop:Ensemble_Noisy}:}} It can be shown by contradiction that
for every $\epsilon>0$ there exists a $\mu_{\epsilon}$ such that
for any $\mu>\mu_{\epsilon}$ it holds that
$\|{\bbB}_e-{\bbC}_e^T\|_1\leq \epsilon/2$. Given that
$\bbSigma_{x}=\bbU_x\bbD_x\bbU_x^T$, and since
$\|{\bbB}_{e}-{\bbC}^T_e\|_1<\epsilon/2$ for $\mu>\mu_{\epsilon}$,
the minimization problem in \eqref{Eq:Ens_L1_PCA} can be
equivalently rewritten as
\begin{equation}\label{Eq:L1_SPCA_w_epsilon}
{\bbC}_e(\epsilon)\in
\arg\min_{\bbC}\textrm{tr}\left(\bbD_{x}^{1/2}
\bbU_x^T(\bbI_{q}-2\bbC^T\bbC+\bbC^T\bbC\bbC^T\bbC)\bbU_x\bbD_x^{1/2}
\right)+\sum_{\rho=1}^{q}2\lambda\|\bbC_{\rho:}^T\|_1+\phi(\bbC,\epsilon,\mu)
\end{equation}
where $\phi(\bbC,\epsilon,\mu)$ is a continuous function of $\bbC$
and $\epsilon$, while $\phi(\bbC,0,\mu)=0$.

Let us now consider how the support of each of the rows of
${\bbC}_e$ is related
 to the support of
the principal eigenvectors $\{\bbu_{s,\rho}\}_{\rho=1}^{r}$. To
this end, remove $\phi(\bbC,\epsilon,\mu)$ from
\eqref{Eq:L1_SPCA_w_epsilon} and consider the minimization problem
\begin{equation}\label{Eq:Only_C_SPCA}
\check{\bbC}_e \in
\arg\min_{\bbC}\textrm{tr}[(\bbI-\bbC^T\bbC)\bbSigma_x(\bbI-\bbC^T\bbC)^T]
+\sum_{\rho=1}^{q}2\lambda\|\bbC_{\rho:}^T\|_1.
\end{equation}

Since the cost in \eqref{Eq:L1_SPCA_w_epsilon}  is continuous, one
recognizes after applying a continuity argument \cite[pg.
15]{Sensitivity_Book}, that for any $\delta>0$  a sufficiently
large $\mu_{\delta}$ can be found such that for any
$\mu>\max(\mu_{\delta},\mu_{\epsilon})$ there exists an optimal
solution ${\bbB}_{e}, {\bbC}_{e}$ in \eqref{Eq:L1_SPCA_w_epsilon},
as well as an optimal solution $\check{\bbC}_e$ in
\eqref{Eq:Only_C_SPCA} such that
$\|{\bbC}_{e}-\check{\bbC}_{e}\|_{1}\leq \delta/2$ and
$\|{\bbB}_{e}-\check{\bbC}_{e}^T\|_1\leq
\|{\bbB}_{e}-{\bbC}_{e}^T\|_1
+\|{\bbC}_{e}-\check{\bbC}_{e}\|_1\leq \delta$ (details are
omitted due to space limitations). As the optimal solutions of
\eqref{Eq:L1_SPCA_w_epsilon} and \eqref{Eq:Only_C_SPCA} can be
arbitrarily close, one considers the simpler of the two in
\eqref{Eq:Only_C_SPCA}.

\textcolor{black}{\noindent Given that
$\bbSigma_x=\bbSigma_s+\bbSigma_w=\bbU_{s,r}\bbD_{s,r}\bbU_{s,r}^T+\bbU_w\bbD_w\bbU_w^T$,
the minimization in \eqref{Eq:Only_C_SPCA} can be rewritten as}
\textcolor{black}{
\begin{align}\label{Eq:Only_C_SPCA_b}
\tilde{\bbC}_{e}\in\arg\min_{\tilde{\bbC}}&\;\;\textrm{tr}\left(
\tilde{\bbC}(\bbSigma_{s,P}+\bbSigma_{w,P})\tilde{\bbC}^T
(\tilde{\bbC}\tilde{\bbC}^T-2\mathbf{I}_{q})\right)
+2\lambda\sum_{\rho=1}^{q}\|\tilde{\bbC}_{\rho:}^T\|_1
\end{align}}
\textcolor{black}{\noindent where $\tilde{\bbC}=\bbC\bbP^T$,
$\bbSigma_{s,P}:=\bbP\bbSigma_s\bbP^T$,
$\bbSigma_{w,P}:=\bbP\bbSigma_w\bbP^T$ while $\bbP$ is a
permutation matrix constructed so that $\bbSigma_{s,P}$ is block
diagonal, and $\tilde{\bbC}_e=\check{\bbC}_{e}\bbP^T$ denotes one
of the optimal solutions of \eqref{Eq:Only_C_SPCA_b}. Since the
$\ell_1$-norm is permutation invariant, it holds that
$\|\bbC_{\rho:}^T\|_1=\|\tilde{\bbC}_{\rho:}^T\|_1$.}

The minimization problem in \eqref{Eq:Only_C_SPCA_b} can be equivalently written as
\begin{align}\label{Eq:Only_C_SPCA_c}
\tilde{\bbC}_{e}\in\arg\min_{\tilde{\bbC}}&\;\;\textrm{tr}\left(
\tilde{\bbC}\bbSigma_{x,P}\tilde{\bbC}^T
(\tilde{\bbC}\tilde{\bbC}^T-2\mathbf{I}_{q})\right)
+2\lambda\sum_{\rho=1}^{q}\sum_{j=1}^{p}\bbT(\rho,j),\textrm{  s. to }|\tilde{\bbC}(\rho,j)|\leq \bbT(\rho,j).
\end{align}
Let $\bbT$ denote the $q\times p$ matrix whose $(\rho,j)$th entry is equal to $\bbT(\rho,j)$. Then, the Lagrangian of \eqref{Eq:Only_C_SPCA_c} is
\begin{equation}\label{Lagrangian_Function}
{\cal L}(\tilde{\bbC},\bbT,\bbL_1,\bbL_2)=\textrm{tr}(\tilde{\bbC}\bbSigma_{x,P}\tilde{\bbC}^T
(\tilde{\bbC}\tilde{\bbC}^T-2\mathbf{I}_{q}))+2\lambda\mathbf{1}_{q\times 1}^T\bbT\mathbf{1}_{p\times 1}+\textrm{tr}(\bbL_1^T(\tilde{\bbC}-\bbT))+\textrm{tr}(\bbL_2^T(-\tilde{\bbC}-\bbT)),
\end{equation}
where $\bbL_1,\bbL_2\in\mathbb{R}^{q\times p}$ and their $(\rho,j)$th entry containts the Lagrange multiplier associated with the constraints $\tilde{\bbC}(\rho,j)\leq \bbT(\rho,j)$ and $-\tilde{\bbC}(\rho,j)\leq \bbT(\rho,j)$, respectively. The first-order optimality conditions imply that the gradient of ${\cal L}(\cdot)$ wrt $\tilde{\bbC}$ should be equal to zero when evaluated at $\tilde{\bbC}_{e}$, i.e.,
\begin{equation}\label{Lagrangian_Grad_C}
2\tilde{\bbC_{e}}\tilde{\bbC_{e}}^T\tilde{\bbC}_{e}\bbSigma_{x,P}+
2\tilde{\bbC}_{e}\bbSigma_{x,P}\tilde{\bbC_{e}}^T\tilde{\bbC}_{e}
-4\tilde{\bbC}_{e}\bbSigma_{x,P}+\bbL_1-\bbL_2=\mathbf{0}_{q\times p}.
\end{equation}
Similarly, the gradient of ${\cal L}(\cdot)$ wrt $\bbT$ should be equal to zero at the optimum solution, $\bbT^{*}$, which leads to
\begin{equation}\label{Lagrangian_Grad_T}
\bbL_1^{*}+\bbL^{*}_2=2\lambda\mathbf{1}_{q\times 1}\mathbf{1}^T_{p\times 1}.
\end{equation}
Moreover, the optimal multipliers should be nonnegative, i.e., $\bbL_1^{*}(\rho,j)\geq 0$ and $\bbL_2^{*}(\rho,j)\geq 0$ for $j,\rho=1,\ldots,q$, while the complementary slackness conditions give that $\bbL_1^{*}(\rho,j)(\tilde{\bbC}_{e}(\rho,j)-T^{*}(\rho,j))=0$ and  $\bbL_2^{*}(\rho,j)(-\tilde{\bbC}_{e}(\rho,j)-T^{*}(\rho,j))=0$
(see e.g., \cite[pg. 316]{Bertsekas_Nonlinear_Book}).
\newline
Let $\bbe_{\rho}\in\mathbb{R}^{q\times 1}$ denote the canonical vector which has a single nonzero entry equal to one at the $\rho$-th position. After multiplying the left hand side (lhs) of \eqref{Lagrangian_Grad_C} from the left with $\bbe_{\rho:}^T$ and from the right with $\tilde{\bbC}_{e,\rho:}$ we obtain
\begin{align}\label{Opt_Cond_1}
2\tilde{\bbC}_{e,\rho:}^T\tilde{\bbC}_{e}^T\tilde{\bbC}_e\bbSigma_{x,P}\tilde{\bbC}_{e,\rho:}
+2\tilde{\bbC}_{e,\rho:}^T\bbSigma_{x,P}\tilde{\bbC}_{e}^T\tilde{\bbC}_e\tilde{\bbC}_{e,\rho:}
&-4\tilde{\bbC}_{e,\rho:}^T\bbSigma_{x,P}\tilde{\bbC}_{e,\rho:}\\
&\hspace{2cm}+\sum_{j=1}^{p}(\bbL_1^{*}(\rho,j)-\bbL_2^{*}(\rho,j))\tilde{\bbC}_{e}(\rho,j)=0\nonumber.
\end{align}
Note that the last summand in \eqref{Opt_Cond_1} is equal to $2\lambda\|\tilde{\bbC}_{\rho:}\|_{1}$. This follows from the aforementioned slackness conditions. Specifically, if $\tilde{\bbC}_{e}(\rho,j)>0$ then $\tilde{\bbC}_{e}(\rho,j)=\bbT^{*}({\rho,j})>0$, which further implies that $\bbL_{2}^{*}(\rho,j)=0$ and from \eqref{Lagrangian_Grad_T} it follows that $\bbL_1^{*}(\rho,j)=2\lambda$. In the same way if
$\tilde{\bbC}_{e}(\rho,j)<0$, then $\tilde{\bbC}_{e}(\rho,j)=-\bbT^{*}({\rho,j})<0$ from which it follows that $\bbL_1^{*}(\rho,j)=0$, thus from \eqref{Lagrangian_Grad_T} we conclude that $\bbL_2^{*}(\rho,j)=2\lambda$. Thus, $(\bbL_1^{*}(\rho,j)-\bbL_2^{*}(\rho,j))\tilde{\bbC}_{e}(\rho,j)=2\lambda|\tilde{\bbC}_{e}(\rho,j)|$, and after some algebraic manipulations on \eqref{Opt_Cond_1} it follows
\begin{equation}\label{SPCA_Constraint_1}
\tilde{\bbC}_{e,\rho:}^T\bbSigma_{x,P}(\bbI_{p\times p}-\tilde{\bbC}_{e}^T\tilde{\bbC}_{e})\tilde{\bbC}_{e,\rho:}=0.5\lambda\|\tilde{\bbC}_{e,\rho:}\|_{1},\;\;\rho=1,\ldots,q.
\end{equation}
Summing the $q$ different equalities in \eqref{SPCA_Constraint_1} we obtain
\begin{equation}\label{SPCA_Constraint_3}
\textrm{tr}(\bbSigma_{x,P}(\bbI_{p\times p}-\tilde{\bbC}_e^T\tilde{\bbC}_{e})\tilde{\bbC}_e^T\tilde{\bbC}_e)=0.5\lambda\sum_{\rho=1}^{q}
\|\tilde{\bbC}_{e,\rho:}\|_{1}
\end{equation}

Equality \eqref{SPCA_Constraint_3} can be used to reformulate the cost in \eqref{Eq:Only_C_SPCA_b} without affecting the optimal solution. Specifically, the cost in \eqref{Eq:Only_C_SPCA_b} can be rewritten as $\textrm{tr}(\tilde{\bbC}\bbSigma_{x,P}\tilde{\bbC}^T\tilde{\bbC}\tilde{\bbC}^T)-
2\textrm{tr}(\tilde{\bbC}\bbSigma_{x,P}\tilde{\bbC}^T)+2\lambda\sum_{\rho=1}^{q}\|\tilde{\bbC}_{\rho:}^T\|_{1}=-\textrm{tr}(\tilde{\bbC}\bbSigma_{x,P}\tilde{\bbC}^T)+1.5\lambda\sum_{\rho=1}^{q}\|\tilde{\bbC}_{\rho,:}^T\|_1$. Using the latter cost expression and expanding the lhs of the $q$ different equality constraints in \eqref{SPCA_Constraint_1} the minimization problem in \eqref{Eq:Only_C_SPCA_b}, is equivalent to
\begin{align}\label{Eq:Only_C_SPCA_Easy_Format}
&\tilde{\bbC}_{e}\in\arg\min
-\sum_{\rho=1}^{q}\tilde{\bbC}_{\rho:}^T\bbSigma_{x,P}\tilde{\bbC}_{\rho:}
+1.5\lambda\sum_{\rho=1}^{q}\|\tilde{\bbC}_{\rho,:}^T\|_1\\
&\textrm{s. to }\left(1-\|\tilde{\bbC}_{\rho:}\|_2^2\right)
\tilde{\bbC}_{\rho:}^T\bbSigma_{x,P}\tilde{\bbC}_{\rho:}
-\sum_{j=1,j\neq \rho}^{q}\left(\tilde{\bbC}_{j:}^T\tilde{\bbC}_{\rho:}\right)
\left(\tilde{\bbC}_{\rho:}^T\bbSigma_{x,P}\tilde{\bbC}_{j:}\right)=0.5\lambda\|\tilde{\bbC}_{\rho:}\|_1,\;{\rho=1},\ldots,{q}.\nonumber
\end{align}
Each one of the summands of the sum in the lhs of the equality constaints in \eqref{Eq:Only_C_SPCA_Easy_Format} can be rewritten as
\begin{equation}\label{Eq:Positive_terms}
\tilde{\bbC}_{j:}^T\tilde{\bbC}_{\rho:}
\tilde{\bbC}_{\rho:}^T\bbSigma_{x,P}\tilde{\bbC}_{j:}
=\tilde{\bbC}_{\rho:}^T\left(\bbSigma_{x,P}\tilde{\bbC}_{j:}\tilde{\bbC}_{j:}^T\right)\tilde{\bbC}_{\rho:}\geq 0, j,\rho=1,\ldots,q\textrm{ and }\rho\neq j.
\end{equation}
Notice that the quantity in \eqref{Eq:Positive_terms} is nonnegative since $\textrm{rank}(\bbSigma_{x,P}\tilde{\bbC}_{j:}\tilde{\bbC}_{j:}^T)=1$, while the single nonzero eigenvalue of $\bbSigma_{x,P}\tilde{\bbC}_{j:}\tilde{\bbC}_{j:}^T$ is  $d_{\max}(\bbSigma_{x,P}\tilde{\bbC}_{j:}\tilde{\bbC}_{j:}^T)=\textrm{tr}(\bbSigma_{x,P}\tilde{\bbC}_{j:}\tilde{\bbC}_{j:}^T)=\tilde{\bbC}_{j:}^T\bbSigma_{x,P}\tilde{\bbC}_{j:}>0$ for $\tilde{\bbC}_{j:}\neq 0$. From the constraints in \eqref{Eq:Only_C_SPCA_Easy_Format} and \eqref{Eq:Positive_terms}, it follows that $\|\tilde{\bbC}_{\rho:}\|_2\leq 1$, otherwise $\|\tilde{\bbC}_{\rho:}\|_{1}$ would be negative resulting a contradiction.

For the time being let us ignore the noise
covariance matrix $\bbSigma_{w,P}$ by setting it to zero, thus $\bbSigma_{x,P}=\bbSigma_{s,P}$.
For the selected sparsity-controlling coefficient
$\lambda$ in \eqref{Eq:Only_C_SPCA_b} assume that the optimal solution has $\|\tilde{\bbC}_{e,\rho:}^T\|_{0}=l_{\rho}$ and
$\|\tilde{\bbC}_{e,\rho:}^T\|_{1}=\kappa_{\rho}$, while
$2 \leq l_{\rho}\leq G$ for $\rho=1,\ldots,q$. This is possible since the $\ell_1$-norm is used
in S-PCA. The case $q=1$ is considered first to demonstrate the main result which is then generalized for $q>1$. Toward this end, let
$\tilde{\bbC}_{1:}=\|\tilde{\bbC}_{1:}\|_{2}^2\cdot\mathbf{u}_{\tilde{c},:1}$, where $\|\mathbf{u}_{\tilde{c},:1}\|_{2}=1$. Moreover, to simplify notation let $c_1=\|\tilde{\bbC}_{1:}\|_{2}^2$, and $\gamma_1=\mathbf{u}_{\tilde{c},:1}^T\bbSigma_{s,P}\mathbf{u}_{\tilde{c},:1}$; thus
$\tilde{\bbC}_{1:}^T\bbSigma_{s,P}\tilde{\bbC}_{1:}=c_1^2\gamma_1\geq 0$. Let $d_1^{*}$ denote the
maximum spectral radius among all possible  $l_1\times l_1$ submatrices of $\bbSigma_{s,P}$ that are formed after keeping $l_{1}$ of its rows and columns with common indices that are determined by the indices of the $l_{1}$ nonzero entries in the optimal $\tilde{\bbC}_{e,1:}=\|\tilde{\bbC}_{e,1:}\|_{2}^2\mathbf{u}_{\tilde{c},e,:1}$, and $\bbu_{\tilde{c},e,:1}$ the optimal selection for $\bbu_{\tilde{c},:1}$. Then, it holds that $\gamma_1=\mathbf{u}_{\tilde{c},:1}^T\bbSigma_{s,P}\mathbf{u}_{\tilde{c},:1}\leq d_1^{*}$ for any unit-vector $\mathbf{u}_{\tilde{c},:1}$ for which $\|\mathbf{u}_{\tilde{c},:1}\|_{0}=l_1$.
With this notation in mind, $q=1$ and $\|\tilde{\bbC}_{e,1:}^T\|_{1}=\kappa_1$ \eqref{Eq:Only_C_SPCA_Easy_Format} is equivalent to
\begin{equation}\label{Eq:Opt_q_1a}
\min _{c_1,\gamma_1}-c^2_1\gamma_1,\;\;\;
\textrm{s. to }(1-c_1^2)c_1^2\gamma_1=0.5\lambda\kappa_1,\;\;0\leq c_1\leq\kappa_1'=\min(1,\kappa_1),\;\;0\leq \gamma_1\leq d_1^{*},
\end{equation}
where the first inequality constraint in \eqref{Eq:Opt_q_1a} follows from the fact that $c_1\leq 1$ and
$c_1=\|\tilde{\bbC}_{1:}\|_2<\kappa_1=\|\tilde{\bbC}_{1:}\|_1$. The Lagrangian of \eqref{Eq:Opt_q_1a} is given as
\begin{equation}\label{Eq:Lagrangian_1}
{\cal L}_1(c_1,\gamma_1,\bbv)=-c^2_1\gamma_1+v_1^{a}[(1-c_1^2)c_1^2\gamma_1-0.5\lambda\kappa_1]
+v_1^{b}[c_1-\kappa_1']-v_1^{c}c_1-v_1^{d}\gamma_1+v_{1}^{e}(\gamma_1-d_1^{*}),
\end{equation}
where $\bbv:=[v_1^{a}\; v_1^{b}\; v_1^{c}\; v_1^{d}\; v_1^{e}]^T$ contains the Lagrange multipliers.
After i) differentiating \eqref{Eq:Lagrangian_1} with respect to $c_1$ and $\gamma_1$; ii) setting the corresponding derivatives equal to zero; and iii) applying the complementary slackness conditions [see also Karush-Kuhn-Tucker necessary optimality condition in \cite[pg. 316]{Bertsekas_Nonlinear_Book}] it follows that the optimal value $v_{1}^{e,*}$ of the multiplier $v_1^{e}$ should be strictly positive. The slackness conditions imply that $v_{1}^{e,*}(\gamma_1^{*}-d_1^{*})=0$, then it follows that at the minimum of \eqref{Eq:Opt_q_1a}
it holds that $\gamma_1^{*}=d_1^{*}$. Now recall that $\gamma_1=\mathbf{u}_{\tilde{c},:1}^T\bbSigma_{s,P}\mathbf{u}_{\tilde{c},:1}$, thus $\gamma_1^{*}$ is formed when $\mathbf{u}_{\tilde{c},:1}=\mathbf{u}_{\tilde{c},e,:1}$ (the optimal direction toward which the optimal row $\bbC_{e,\rho:}$ is pointing).

Recall that $d_1^{*}=\max_{\bbu_{\tilde{c},:1}} \mathbf{u}_{\tilde{c},:1}^T\bbSigma_{s,P}\mathbf{u}_{\tilde{c},:1}$ subject to $\|\mathbf{u}_{\tilde{c},:1}\|_{2}=1$ and $\|\mathbf{u}_{\tilde{c},:1}\|_{0}=l_1$.
Next, we demonstrate that if $\mathbf{u}_{\tilde{c},e,:1}^T\bbSigma_{s,P}\mathbf{u}_{\tilde{c},e,:1}=d_1^{*}$, then
there exists a
column, say the $i_{1}$th in $\tilde{\bbU}_{s,r}$, with support
$\tilde{{\cal S}}_{i_{1}}$ such that
$\|\bbu_{\tilde{c},e,:1}^T(\bar{\tilde{{\cal S}}}_{i_{1}})\|_{1}=0$,
while $\|\bbu_{\tilde{c},e,:1}^T(\tilde{\cal S}_{i_{1}})\|_{1}
>0$ and $\bar{\tilde{{\cal S}}}_{i_{1}}$ denotes the complement of
$\tilde{{\cal S}}_{i_{1}}$. Since $\tilde{\bbC}_{e,1:}^T$ is a scaled version of
$\bbu_{\tilde{c},e,:1}^T$, the latter property will further
imply that $\|\tilde{\bbC}_{e,1:}^T(\bar{\tilde{{\cal
S}}}_{i_{1}})\|_{1}=0$, while $\|\tilde{\bbC}_{e,1:}^T(\tilde{\cal
S}_{i_{1}})\|_{1}\geq\xi_{1}'(\lambda)$, where
$\xi_{1}'(\lambda)$ is
strictly positive. Equivalently, we will show that
${\cal I}_{1}:={\cal S}(\tilde{\bbC}_{e,1:})={\cal S}({\bbu}_{\tilde{c},e,:1})\subseteq
{\cal G}_{k_{1}}$, where ${\cal
G}_{k_{1}}=\tilde{{\cal S}}_{i_{1}}$ corresponds to the index set of
the entries of $\bbs_P=\bbP\bbs$ that belong to, say the
$k_{1}$th diagonal block of $\bbSigma_{s,P}$ and
$k_{1}\in\{1,\ldots,K\}$. To this end, let $\bbu_{\tilde{c},:1}^T=[(\bbu_{\tilde{c},:1}^1)^T,$
$(\bbu_{\tilde{c},:1}^2)^T,\ldots,(\bbu_{\tilde{c},:1}^K)^T]$,
where each subvector $\bbu_{\tilde{c},:1}^k$ has $G$ entries; and let ${\cal I}_{1,k}:={\cal
S}(\bbu_{\tilde{c},:1}^k)$ with $\sum_{k=1}^{K}|{\cal
I}_{1,k}|=l_{1}$. Then, it follows that
\begin{equation}\label{Eq:u_Sigma_u_Upper_bound}
\mathbf{u}_{\tilde{c},:1}^T\bbSigma_{s,P}\mathbf{u}_{\tilde{c},:1}
=\sum_{k=1}^{K}(\bbu_{\tilde{c},:1}^k)^T\bbSigma_{s_{{\cal G}_k}}\bbu_{\tilde{c},:1}^k
\leq \sum_{k=1}^{K}d_{\max}(\bbSigma_{s_{{\cal G}_k}}^{l_1})\|{\bbu_{\tilde{c},:1}^k}\|_{2}^2,
\end{equation}
where $d_{\max}(\bbSigma_{s_{{\cal G}_k}}^{l_1})$ denotes the spectral radius of the $l_1\times l_1$
submatrix $\bbSigma_{s_{{\cal G}_k}}^{l_1}$ formed
by the $G\times G$ diagonal block  $\bbSigma_{s_{{\cal G}_k}}$ of $\bbSigma_{x,P}$ after keeping
$l_1$ of its rows and columns with common indices. The inequality in \eqref{Eq:u_Sigma_u_Upper_bound} follows since each subvector $\bbu_{\tilde{c},:1}^k$ of $\bbu_{\tilde{c},:1}$ can
have at most $l_1$ nonzero entries. If $d_1^{l_1}$ denotes the maximum spectral radius that can be achieved by any $l_1\times l_1$ submatrix $\bbSigma_{s_{{\cal }_k}}^{l_1}$ that is contained in a diagonal block $\bbSigma_{s_{{\cal G}_k}}$ of $\bbSigma_{s,P}$ for $k=1,\ldots,K$, then from \eqref{Eq:u_Sigma_u_Upper_bound} and since $\sum_{k=1}^{K}\|\bbu_{\tilde{c},:1}^k\|_{2}^2=1$ it holds that $\mathbf{u}_{\tilde{c},:1}^T\bbSigma_{s,P}\mathbf{u}_{\tilde{c},:1}\leq d_1^{l_1}$. Thus, it should hold that $d_1^{l_1}=d_1^{*}$. Then, the max value $d_1^{*}$ can be attained if and only if the indices of the nonzero entries of $\mathbf{u}_{\tilde{c},e,:1}$ satisfy ${\cal I}_1\subseteq {\cal G}_{k_1}$ for a $k_1\in\{1,\ldots,K\}$.  This further implies that there
exists an eigenvector
$\tilde{\bbu}_{s,i_1}:=\bbP{\bbu}_{s,i_1}$ with support
$\tilde{{\cal S}}_{i_1}={\cal G}_{k_1}$, for which ${{\cal
I}_{1}}\subseteq \tilde{{\cal S}}_{i_1}$. Thus, it is deduced that
$\bbu_{\tilde{c},e,:1}^T(\bar{\tilde{{\cal S}}}_{i_1})=0$
and $\|\bbu_{\tilde{c},e,:1}^T(\tilde{\cal
S}_{i_1})\|_{1}\geq \xi_{1}'(\lambda)>0$ since the
$l_1$ nonzero entries of $\bbu_{\tilde{c},e,:1}^T$ have
indices in $\tilde{{\cal S}}_{i_1}$. Positivity of
$\xi_{1}'(\lambda)$ is ensured since
$\|\bbu_{\tilde{c},e,:1}^T\|_2=1$ and $\lambda$ is selected such that $\tilde{\bbC}_{e,1:}=(c_1^{*})^2\bbu_{\tilde{c},e,:\rho}\neq \mathbf{0}$.
Since
$\check{\bbC}_{e}=\tilde{\bbC}_{e}\bbP$ in \eqref{Eq:Only_C_SPCA}
results from permuting the columns of $\tilde{\bbC}_{e}$, it
follows that $\|\check{\bbC}_{e,1:}^T(\bar{{{\cal
S}}}_{i_{\rho}})\|_{1}=0$ and $\|\check{\bbC}_{e,1:}^T({\cal
S}_{i_{\rho}})\|_{1}\geq \xi_{1}'(\lambda)>0$, where
${\cal S}_{i_\rho}={\cal S}(\bbu_{s,i_{\rho}})$.

We generalize the previous claim for the case when $q>1$. As before we reexpress each of the rows of
$\tilde{\bbC}$ as $\tilde{\bbC}_{\rho:}=\|\tilde{\bbC}_{\rho:}\|_{2}^2\bbu_{\tilde{c},:\rho}$, with
$\|\bbu_{\tilde{c},:\rho}\|_2=1$ for $\rho=1,\ldots,q$. No other assumptions are imposed for the direction vectors
$\bbu_{\tilde{c},:\rho}$. Further, let $c_{\rho}=\|\tilde{\bbC}_{\rho:}\|_{2}$ and
$\gamma_{\rho}=\bbu_{\tilde{c},:\rho}^T\bbSigma_{s,P}\bbu_{\tilde{c},:\rho}$, where $\rho=1,\ldots,q$.
Moreover, let $\delta_{j,\rho}=(\bbu_{\tilde{c},:j}^T\bbu_{\tilde{c},:\rho})(\bbu_{\tilde{c},:j}^T\bbSigma_{s,P}\bbu_{\tilde{c},:\rho})$ for $j\neq \rho$ and further notice that $\delta_{j,\rho}=\delta_{\rho,j}\geq 0$ [cf. \eqref{Eq:Positive_terms}], while
$\delta_{j,\rho}\leq \gamma_{\rho}$ and $\delta_{j,\rho}\leq \gamma_{j}$. Also recall that $\lambda$ has been selected such that $\{\|\tilde{\bbC}_{e,\rho:}\|_{1}=\kappa_{\rho}\}_{\rho=1}^{q}$ and $\{\|\tilde{\bbC}_{e,\rho:}\|_{0}=l_{\rho}\}_{\rho=1}^{q}$ where $2\leq l_{\rho}\leq G$. Then, the minimization problem in \eqref{Eq:Only_C_SPCA_Easy_Format} can be equivalently rewritten as
\begin{align}\label{Eq:Opt_q_greater_1}
\min-\sum_{\rho=1}^{q}c_{\rho}^2\gamma_{\rho},\;\;\;
&\textrm{s. to }
(1-c_{\rho}^2)c_{\rho}^2\gamma_{\rho}-\sum_{j=1,j\neq \rho}^{q}c_{\rho}^2c_{j}^2\delta_{j,\rho}=0.5\lambda\kappa_{\rho},\;,
0\leq c_{\rho}\leq \kappa_{\rho}':=\min(1,\kappa_{\rho})\;\nonumber\\
& \hspace{0cm}0\leq \gamma_{\rho}\leq d_{\rho}^{*},\;\;,0\leq \delta_{j,\rho}\leq \gamma_{\rho},\;\;\delta_{j,\rho}\leq \gamma_{j},\;\;\delta_{j,\rho}=\delta_{\rho,j},\;j\neq \rho,\;
j,\rho=1,\ldots,q
\end{align}
where $d_{\rho}^{*}$ corresponds to the maximum value that $\bbu_{\tilde{c},\rho:}^T\bbSigma_{x,P}\bbu_{\tilde{c},\rho:}$ can attain when $\|\tilde{\bbC}_{\rho:}\|_{0}=l_{\rho}$, while $\bbu_{\tilde{c},\rho:}$ are selected such that the constraints in \eqref{Eq:Opt_q_greater_1} are satisfied. The Lagrangian function of \eqref{Eq:Opt_q_greater_1} is given as
\begin{align}\label{Eq:Lagrangian_2}
{\cal L}_2(\{c_{\rho},\gamma_{\rho}\}_{\rho=1}^{q},\{\delta_{j,\rho}\},
\bbv)=&
-\sum_{\rho=1}^{q}c_{\rho}^2\gamma_{\rho}+\sum_{\rho=1}^{q}v_{\rho}^{a}
[(1-c_{\rho}^2)c_{\rho}^2\gamma_{\rho}-\sum_{j=1,j\neq \rho}^{q}c_{\rho}^2c_{j}^2\delta_{j,\rho}-0.5\lambda\kappa_{\rho}]\\
&+\sum_{\rho=1}^{q}\left[v_{\rho}^{b}(c_{\rho}-\kappa_{\rho}')-v_{\rho}^{c}c_{\rho}
+v_{\rho}^{d}(\gamma_{\rho}-d_{\rho}^{*})-v_{\rho}^{e}\gamma_{\rho}\right]\nonumber\\
&+
\sum_{\rho=1}^{q}\sum_{j=1,j\neq \rho}^{q}\left[v_{j,\rho}^{f}(\delta_{j,\rho}-\gamma_{\rho})
+v_{j,\rho}^{g}(\delta_{j,\rho}-\gamma_{j})+v_{j,\rho}^{h}(\delta_{j,\rho}-\delta_{\rho,j})-v_{j,\rho}^{i}\delta_{j,\rho}\right],\nonumber
\end{align}
where $\bbv$ is a vector that contains the Lagrange multipliers $v_{\rho}^a,v_{\rho}^b,v_{\rho}^c,v_{\rho}^d,v_{\rho}^{e},v_{j,\rho}^f,v_{j,\rho}^{g}$, $v_{j,\rho}^h$ and $v_{j,\rho}^i$.
The KKT conditions are applied next to derive necessary conditions that the optimal solution of
\eqref{Eq:Opt_q_greater_1} should satisfy. This involves i) differentiating \eqref{Eq:Lagrangian_2} wrt $c_{\rho}$, $\gamma_{\rho}$ and $\delta_{j,\rho}$; ii) setting the corresponding derivatives equal to zero; and iii) applying the complementary slackness conditions for the optimal multipliers $\bbv^{*}$. Then, it follows that at the minimum of \eqref{Eq:Opt_q_greater_1} it should hold that
$\delta_{j,\rho}^{*}=0$, and $\gamma_{\rho}^{*}=d_{\rho}^{*}$ for $j,\rho=1,\ldots,q$ and $j\neq \rho$. From the definition of $\delta_{j,\rho}$ it follows that $\delta_{j,\rho}^{*}$ is formed using the optimal vectors $\bbu_{\tilde{c},e,\rho:}$, i.e.,
$\delta_{j,\rho}^{*}=(\bbu_{\tilde{c},e,:j}^T\bbu_{\tilde{c},e,:\rho})(\bbu_{\tilde{c},e,:j}^T\bbSigma_{s,P}\bbu_{\tilde{c},e,:\rho})$. Since $\delta_{j,\rho}^{*}=0$, it follows that the optimal direction vector $\bbu_{\tilde{c},e,:\rho}$ should be selected in \eqref{Eq:Only_C_SPCA_Easy_Format} such that $\bbu_{\tilde{c},e,:j}^T\bbu_{\tilde{c},e,:\rho}=0$,
or $\bbu_{\tilde{c},e,:j}^T\bbSigma_{s,P}\bbu_{\tilde{c},e,:\rho}=0$ for $j\neq \rho$, while $\gamma_{\rho}^{*}=\bbu_{\tilde{c},e,:\rho}^T\bbSigma_{s,P}\bbu_{\tilde{c},e,:\rho}$ iis equal to the maximum possible value $d_{\rho}^{*}$. Since $\tilde{\bbC}_{e,\rho:}=(c_{\rho}^{*})^2\bbu_{\tilde{c},e,:\rho}$ it follows that the rows of the optimal matrix $\tilde{\bbC}_{e}$ should be selected such that
either they are orthogonal $\tilde{\bbC}_{e,\rho:}^T\tilde{\bbC}_{e,j:}=0$, or
$\tilde{\bbC}_{e,\rho:}^T\bbSigma_{s,P}\tilde{\bbC}_{e,j:}=0$.
In summary the direction vector for the $\rho$th row of the optimal matrix $\tilde{\bbC}_{e}$ in \eqref{Eq:Only_C_SPCA_Easy_Format}, namely $\bbu_{\tilde{c},e,:\rho}$, should be selected such that
\begin{equation}\label{Eq:u_q_greater_1}
\bbu_{\tilde{c},e,:\rho}=\arg\max_{\bbu_{\tilde{c},\rho:}}\bbu_{\tilde{c},:\rho}^T\bbSigma_{s,P}
\bbu_{\tilde{c},:\rho},\;\;\;\textrm{s. to }(\bbu_{\tilde{c},:j}^T\bbu_{\tilde{c},e,:\rho})(\bbu_{\tilde{c},:j}^T\bbSigma_{s,P}\bbu_{\tilde{c},:\rho})=0,\;\;\|\bbu_{\tilde{c},:\rho}\|_2=1,\;\;\|\bbu_{\tilde{c},e,:\rho})\|_{1}=l_{\rho},\nonumber
\end{equation}
where $\rho=1,\ldots,q,\;j\neq \rho.$ Using similar reasoning as in the case where $q=1$ it follows that for every optimal row $\tilde{\bbC}_{e,\rho:}$ there exists $i_{\rho}\in\{1,\ldots,r\}$ such that  ${\cal I_{\rho}}={\cal S}(\tilde{\bbC}_{e,\rho:}^T)\subseteq{\cal S}(\tilde{\bbu}_{s,i_{\rho}})$
Letting
$\bar{\tilde{{\cal S}}}_{i_{\rho}}$ be the complement of
$\tilde{{\cal S}}_{i_{\rho}}$, it is deduced that
$\bbu_{\tilde{c},e,:\rho}^T(\bar{\tilde{{\cal S}}}_{i_{\rho}})=0$
and $\|\bbu_{\tilde{c},e,\rho:}^T(\tilde{\cal
S}_{i_{\rho}})\|_{1}\geq \xi_{\rho}'(\lambda)>0$ since the
$l_{\rho}$ nonzero entries of $\bbu_{\tilde{c},e,:\rho}^T$ have
indices in $\tilde{{\cal S}}_{i_{\rho}}$. Positivity of
$\xi_{\rho}'(\lambda)$ is ensured since
$\|\bbu_{\tilde{c},e,:\rho}^T\|_2=1$ and $\tilde{\bbC}_{e,\rho:}=(c_{\rho}^{*})^2\bbu_{\tilde{c},e,:\rho}\neq \mathbf{0}$ for the selected $\lambda$.
Then, it follows readily that
$\|\tilde{\bbC}_{e,\rho:}(\bar{\tilde{{\cal
S}}}_{i_{\rho}})\|_{1}=0$ while
$\|\tilde{\bbC}_{e,\rho:}(\tilde{\cal S}_{i_{\rho}})\|_{1}\geq
\xi_{\rho}'(\lambda)>0$ for $\rho=1,\ldots,q$. Since
$\check{\bbC}_{e}=\tilde{\bbC}_{e}\bbP$ in \eqref{Eq:Only_C_SPCA}
results from permuting the columns of $\tilde{\bbC}_{e}$ it
follows that $\|\check{\bbC}_{e,\rho:}^T(\bar{{{\cal
S}}}_{i_{\rho}})\|_{1}=0$ and $\|\check{\bbC}_{e,\rho:}^T({\cal
S}_{i_{\rho}})\|_{1}\geq \xi_{\rho}'(\lambda)>0$, where
${\cal S}_{i_\rho}={\cal S}(\bbu_{s,i_{\rho}})$.

\textcolor{black}{The latter property was proved under the
assumption that $\bbSigma_{w,P}=0$. Consider now the general case
where $\bbSigma_{w,P}\neq \mathbf{0}$, thus $\bbSigma_{x,P}\neq \bbSigma_{s,P}$.
An upper bound on the noise variance will be determined that ensures
the validity of the earlier claims about
$\tilde{\bbC}_{e,\rho:}^T$ (or $\check{\bbC}_{e,\rho:}^T$)
established in the noiseless case. Let
$\check{\bbu}_{\tilde{c},:\rho}$ be a direction vector that results a row vector $\check{\bbC}_{\rho:}$ that belongs to the
constraint set of \eqref{Eq:Only_C_SPCA_Easy_Format}, while
$\|\check{\bbu}_{\tilde{c},:\rho}\|_0=\|\check{\bbC}_{\rho:}\|_0=l_{\rho}$. Further, assume
that the support of $\check{\bbu}_{\tilde{c},:\rho}$ is different
from the support of the optimal $\bbu_{\tilde{c},e,:\rho}^T$
evaluated when $\bbSigma_{w,P}=\mathbf{0}$. One
sufficient condition to ensure optimality of
$\{\bbu_{\tilde{c},e,:\rho}^T\}_{\rho=1}^{q}$ in the presence of
noise is that
\begin{equation}\label{Eq:Nec_Condition}
\check{\bbu}_{\tilde{c},:\rho}^T\bbSigma_{s,P}\check{\bbu}_{\tilde{c},:\rho}
+\check{\bbu}_{\tilde{c},:\rho}^T\bbSigma_{w,P}\check{\bbu}_{\tilde{c},:\rho}<d^{*}_{\rho}
=\bbu_{\tilde{c},e,:\rho}^T\bbSigma_{s,P}\bbu_{\tilde{c},e,:\rho},\;\;
\rho=1,\ldots,q.
\end{equation}}
\textcolor{black}{\noindent for any
$\check{\bbu}_{\tilde{c},:\rho}$ that results a feasible $\check{\bbC}_{\rho:}$ in \eqref{Eq:Only_C_SPCA_Easy_Format}, while
$\|\check{\bbu}_{\tilde{c},:\rho}\|_{0}=l_{\rho}$ and ${\cal
S}(\check{\bbu}_{\tilde{c},:\rho})\neq {\cal
S}(\bbu_{\tilde{c},e,:\rho})$. Given that
$\check{\bbu}_{\tilde{c},:\rho}^T\bbSigma_{w,P}\check{\bbu}_{\tilde{c},:\rho}\leq
d_{\textrm{max}}(\bbSigma_{w,P})$, it follows that
\eqref{Eq:Nec_Condition} will be satisfied when} \textcolor{black}
{\begin{equation}\label{Eq:Noise_upper}
d_{\textrm{max}}(\bbSigma_{w,P})<d^{*}_{\rho}
-\check{\bbu}_{\tilde{c},:\rho}^T\bbSigma_{s,P}
\check{\bbu}_{\tilde{c},:\rho}.
\end{equation}}
\textcolor{black}{Note that
$\check{\bbu}_{\tilde{c},:\rho}^T\bbSigma_{s,P}\check{\bbu}_{\tilde{c},:\rho}<
d^{*}_{\rho}$ since $\check{\bbu}_{\tilde{c},:\rho}$ does not have
the same support as $\bbu_{\tilde{c},e,\rho:}^T$ that maximizes
the problem at the bottom of pg. 28, in which $\bbSigma_{w,P}=\mathbf{0}$.
Thus, the quantity in the right hand side of \eqref{Eq:Noise_upper}, denoted as
$\Delta(\bbSigma_s)$, will be positive.}

\textcolor{black}{What remains to establish are the properties
stated in Prop. 1 for $\bbC_{e,\rho:}^T$ and $\bbB_{e,:\rho}$ with
$\rho=1,\ldots,q$. To this end, recall that for any
$\mu>\max(\mu_{\delta},\mu_{\epsilon})$, and for each
$\tilde{\bbC}_{e}$, or equivalently $\check{\bbC}_{e}$, there
exists an optimal solution ${\bbC}_e$ and ${\bbB}_e$ of
\eqref{Eq:Ens_L1_PCA} for which
$\|{\bbC}_{e}-\check{\bbC}_{e}\|_1\leq\delta/2$ and
$\|{\bbB}_{e}-\check{\bbC}_{e}^T\|_1\leq\delta$, where
$\check{\bbC}_{e}=\tilde{\bbC}_e\bbP$. Then,
$\|{\bbC}_{e,\rho:}^T-\check{\bbC}_{e}\|_{1}\leq\delta/2$ for
$\rho=1,\ldots,q$. Then, it readily follows that
$\|{\bbC}_{e,\rho:}^T(\bar{\cal S}_{i_{\rho}})\|_{1}\leq \delta/2$
since $\check{{\bbC}}_{e,\rho:}^T(\bar{\cal
S}_{i_{\rho}})=\mathbf{0}^T$. Moreover, $
\|\check{{\bbC}}_{e,\rho:}^T({\cal S}_{i_{\rho}})\|-\delta/2\leq
\|{\bbC}_{e,\rho:}^T({\cal S}_{i_{\rho}})\|_{1}\leq
\|\check{{\bbC}}_{e,\rho:}^T({\cal S}_{i_{\rho}})\|+\delta/2$.
Notice that the lower bound $ \|\check{{\bbC}}_{e,\rho:}^T({\cal
S}_{i_{\rho}})\|-\delta/2\geq \xi_{\rho}'(\lambda)-\delta/2$
can be made strictly positive by pushing $\delta/2$ arbitrarily
close to zero, which is possible by increasing $\mu$. However,
$\xi_{\rho}(\lambda)$ remains strictly positive for the values of $\lambda_{\rho}$ considered here,
since it does not depend on
$\mu$. These properties can also be established for
$\bbB_{e,:\rho}$ using similar arguments. \myQED}

\noindent \emph{D. Proof of Proposition
\ref{Prop:Asymptotic_Normality}:} In the noiseless case the
training matrix $\bbX_n=\bbS_n$ (note the dependence on $n$) can
be written as
$\bbS_n=\bbU_{s,q}\bbU_{s,q}^T\bbS_n+\bbU_{s,p-q}\bbU_{s,p-q}^T\bbS_n$.
For notational convenience let
$\bbGamma_{q,n}=\bbU_{s,q}^T\bbS_n$, and
$\bbz_{q,n}=\textrm{vec}(\bbU_{s,p-q}$ $\bbU_{s,p-q}^T\bbS_n)=
\left((\bbU_{s,p-q}^T\bbS_n)^T\otimes\bbI_{p}\right)\textrm{vec}(\bbU_{s,p-q})$.
Using vec notation, it holds  that
$\textrm{vec}(\bbS_n)=(\bbGamma_{q,n}^T\otimes \bbI_{p})$
$\textrm{vec}(\bbU_{s,q})+\bbz_{q,n}$. Moreover, let
$\bbb=\textrm{vec}(\bbB)=\textrm{vec}(\bbU_{s,q})+\sqrt{n^{-1}}\tilde{\bbb}$,
where $\sqrt{n^{-1}}\tilde{\bbb}$ quantifies the estimation error
present when estimating $\bbU_{s,q}$ via \eqref{Eq:Btaup1}. Using
this  notation and after applying some algebraic manipulations the
cost in \eqref{Eq:Btaup1} can be reformulated as
\begin{align}\label{Eq:Btaup1_reform}
J_{b}(\tilde{\bbb})&:= \|(\bbGamma_{q,n}^T\otimes
\bbI_{p})\textrm{vec}(\bbU_{s,q})
+\bbz_{q,n}-\left((\hat{\bbC}_{\tau,n}\bbS_n)^T\otimes\bbI_{p}\right)\bbb\|_{2}^2
+\sum_{\rho=1}^{q}\lambda_{\rho,n}\sum_{j_{\rho}=p(\rho-1)+1}^{\rho
p}|\bbb(j_{\rho})|
\nonumber\\&+\mu\|\bbb-\hat{\bbc}_{\tau,n}^t\|_2^2 =
\|-\sqrt{n^{-1}}\left[(\hat{\bbC}_{\tau,n}\bbS_n)^T\otimes
\bbI_{p}\right]\tilde{\bbb}
+\bbz_{q,n}-\sqrt{n^{-1}}\left[({\bbE}_{\tau,n}^c\bbS_n)^T\otimes \bbI_p\right]\textrm{vec}(\bbU_{s,q})\|_2^2+\nonumber\\
&\hspace{0.5cm}
\sum_{\rho=1}^{q}\lambda_{\rho,n}\sum_{j_{\rho}=(\rho-1)p+1}^{\rho
p}
|\sqrt{n^{-1}}\tilde{\bbb}(j_{\rho})+\textrm{vec}(\bbU_{s,q})(j_{\rho})|
+\mu\|\sqrt{n^{-1}}\tilde{\bbb}-\sqrt{n^{-1}}{\bbe}_{\tau,n}^{c}\|_2^2
\end{align}
where the second inequality follows after replacing $\bbb$ with
$\textrm{vec}(\bbU_{s,q})+\sqrt{n^{-1}}\tilde{\bbb}$ in all three
terms in the expression following $J_{b}(\tilde{\bbb})$. Moreover,
$\hat{\bbc}_{\tau,n}^t:=\textrm{vec}(\hat{\bbC}_{\tau,n}^T)$,
${\bbe}_{\tau,n}^{c}:=\textrm{vec}(({\bbE}_{\tau,n}^c)^T)$,
and $\textrm{vec}(\bbU_{s,q})(j)$ denotes the $j$th element of
$\textrm{vec}(\bbU_{s,q})$. Recall that the optimal
solution of \eqref{Eq:Btaup1} is $\hat{\bbB}_{\tau,n}$, and let
${\breve{\bbb}}_{n}:=
\sqrt{n}[\textrm{vec}(\hat{\bbB}_{\tau,n})-\textrm{vec}(\bbU_{s,q})]$.
We will show that the error $\breve{{\bbb}}_{n}$ which minimizes
\eqref{Eq:Btaup1_reform} and corresponds to the estimate
$\hat{\bbB}_{\tau,n}$ converges to a Gaussian random variable,
thus establishing the first result in Prop.
\ref{Prop:Asymptotic_Normality}.

\noindent To this end, consider the cost
$J_{b}(\tilde{\bbb})-J_{b}(\mathbf{0})$ which has the same optimal
solution as $J_{b}(\tilde{\bbb})$, since $J_{b}(\mathbf{0})$ is a
constant. After performing some algebraic manipulations we can
readily obtain
\begin{align}\label{Eq:J_b_nz_J_b_0}
\hspace{-0.2cm}J_{b}(\tilde{\bbb})-J_{b}(\mathbf{0})
=&n^{-1}\tilde{\bbb}^T\left[(\hat{\bbC}_{\tau,n}\bbS_n\bbS_n^T\hat{\bbC}_{\tau,n}^T
+\mu\bbI_{q})\otimes \bbI_{p\times p}\right]\tilde{\bbb}-
2n^{-0.5}\tilde{\bbb}^T\left[(\hat{\bbC}_{\tau,n}\bbS_n\bbS_n^T\bbU_{s,p-q})
\otimes\bbI_{p}\right]\textrm{vec}(\bbU_{s,p-q})\nonumber\\
&-2\mu n^{-1}\tilde{\bbb}^T
{\bbe}_{\tau,n}^c+2n^{-1}\tilde{\bbb}^T
\left[(\hat{\bbC}_{\tau,n}\bbS_n\bbS_n^T({\bbE}_{\tau,n}^c)^T)\otimes\bbI_{p}\right]
\textrm{vec}(\bbU_{s,q})\nonumber\\
&\hspace{-2cm}+n^{-1}\sum_{\rho=1}^{q}\sum_{j_{\rho}=(\rho-1)p+1}^{\rho
p} \sqrt{n}\lambda_{\rho,n}\hat{w}_{j_{\rho},\rho,n}\sqrt{n}
\left[\left|\textrm{vec}(\bbU_{s,q})(j_{\rho})+\sqrt{n^{-1}}\tilde{\bbb}(j_{\rho})\right|-
\left|\textrm{vec}(\bbU_{s,q})(j_{\rho})\right|\right].
\end{align}

Next, it is proved that $J_{b}(\tilde{\bbb})-J_{b}(\mathbf{0})$
converges in distribution to a cost $G_b(\tilde{\bbb})$, whose
minimum will turn out to be the limiting point at which
$\breve{\bbb}_{n}$ converges in distribution as
$n\rightarrow\infty$. It follows from (a1) that
$\hat{\bbSigma}_{s,n}=n^{-1}\bbS_n\bbS_n^T$ converges almost
surely (a.s.) to $\bbSigma_s$ as $n\rightarrow\infty$, whereas
$\hat{\bbC}_{\tau,n}$ converges in distribution to $\bbU_{s,q}^T$
(this follows from the asymptotic normality assumption). Then,
Slutsky's theorem, e.g., see \cite{Probability_Book}, implies that
the first term in \eqref{Eq:J_b_nz_J_b_0} converges in
distribution to
$\tilde{\bbb}^T(\bbD_{s,q}\otimes\bbI_{p})\tilde{\bbb}$. Recalling
that the estimation error ${\bbe}_{\tau,n}^c$ is assumed to
converge to a zero-mean Gaussian distribution with finite
covariance, the third term converges in distribution (and in
probability) to $0$. Taking into account (a1) and  that
$\bbS_n\bbS_n^T=n\hat{\bbSigma}_{s,n}=n\bbSigma_s+{\bbE}_{s,n}$,
where $n^{-1}{\bbE}_{s,n}$ corresponds to the covariance
estimation error, it follows readily that the second term in
\eqref{Eq:J_b_nz_J_b_0} is equal to
\begin{align}\label{Eq:Analyzing_First_Summand}
2n^{-0.5}\tilde{\bbb}^T\left[(\hat{\bbC}_{\tau,n}\bbS\bbS^T\bbU_{s,p-q})
\otimes\bbI_{p}\right]\textrm{vec}(\bbU_{s,p-q})&=
2n^{-0.5}\tilde{\bbb}^T\left[(\bbU_{s,q}^T{\bbE}_{s,n}\bbU_{s,p-q})
\otimes\bbI_{p}\right]\textrm{vec}(\bbU_{s,p-q})\nonumber\\
&\hspace{-7.5cm}
+2\tilde{\bbb}^T\left[({\bbE}_{\tau,n}^c\bbSigma_{s}\bbU_{s,p-q})\otimes
\bbI_{p}\right]\textrm{vec}(\bbU_{s,p-q})
+2n^{-1}\tilde{\bbb}^T\left[({\bbE}_{\tau,n}^c{\bbE}_{s,n}\bbU_{s,p-q})
\otimes \bbI_{p}\right]\textrm{vec}(\bbU_{s,p-q}).
\end{align}
Recall that ${\bbE}_{\tau,n}$ converges in distribution to a
zero-mean Gaussian random variables with finite covariance,
whereas ${\bbE}_{s,n}$ adheres to a Wishart distribution with
scaling matrix $\bbSigma_s$ and $n$ degrees of freedom \cite[pg.
47]{Jolliffe_PCA_Book}. Then, it follows readily that the first
and third terms in \eqref{Eq:Limit_First_Summand} converge to zero
in distribution (thus in probability too). Then, we have that the
lhs of \eqref{Eq:Analyzing_First_Summand}
converges to
\begin{equation}\label{Eq:Limit_First_Summand}
2\tilde{\bbb}\left[({\bbE}_{c}\bbSigma_{s}\bbU_{s,p-q})\otimes
\bbI_{p}\right] \textrm{vec}(\bbU_{s,p-q})
=2\tilde{\bbb}^T\textrm{vec}(\bbU_{s,p-q}\bbU_{s,p-q}^T\bbSigma_s{\bbE}_{c}^T)
\end{equation}
where ${\bbE}_{c}$ denotes the Gaussian random matrix at
which ${\bbE}_{\tau,n}^c$ converges in distribution as
$n\rightarrow\infty$. Similarly, we can  show that the fourth
summand in \eqref{Eq:J_b_nz_J_b_0} converges in distribution, as
$n\rightarrow\infty$, to
\begin{equation}\label{Eq:Limit_Fourth_Summand}
2\tilde{\bbb}^T\left[(\bbU_{s,q}^T
\bbSigma_{s}{\bbE}_{c}^T)\otimes
\bbI_{p}\right]\textrm{vec}(\bbU_{s,q})=
2\tilde{\bbb}^T\textrm{vec}(\bbU_{s,q}{\bbE}_{c}\bbSigma_{s}\bbU_{s,q}).
\end{equation}
The limiting noise terms in \eqref{Eq:Limit_First_Summand} and
\eqref{Eq:Limit_Fourth_Summand} are zero-mean and uncorrelated.
Now, we examine the limiting behavior of the double sum in
\eqref{Eq:J_b_nz_J_b_0}. If $\textrm{vec}(\bbU_{s,q})(j)\neq 0$
then $\lim_{n\rightarrow\infty}\sqrt{n}$
$\left[|\textrm{vec}(\bbU_{s,q})(j)\right.$ $+\sqrt{n^{-1}}
\tilde{\bbb}(j)|-$ $\left.|\textrm{vec}(\bbU_{s,q})(j)|
\right]=\textrm{sgn}[\textrm{vec}(\bbU_{s,q})(j)]\tilde{\bbb}(j)$.
Since $\hat{w}_{j,\rho,n}$ converges in probability to
$|\textrm{vec}(\bbU_{s,q})$ $(j)|^{-\gamma}$, we can deduce  that
if $\lambda_{\rho,n}$ is selected as suggested by the first limit
in \eqref{Eq:Scaling_Law_lambda}, then the corresponding term in
the double sum in \eqref{Eq:J_b_nz_J_b_0} goes to zero in
distribution (and in probability) as $n\rightarrow\infty$.

For the case where $\textrm{vec}(\bbU_{s,q})(j)=0$, it holds that
$\sqrt{n}\left[|\textrm{vec}(\bbU_{s,q})(j)+\sqrt{n^{-1}}
\tilde{\bbb}(j)|-|\textrm{vec}(\bbU_{s,q})(j)|\right]$
$=|\tilde{\bbb}(j)|$, and also
$\sqrt{n}\lambda_{\rho,n}\hat{w}_{j,\rho,n}=
\sqrt{n}\lambda_{\rho,n}n^{\gamma/2}(\sqrt{n}|\textrm{vec}
(\hat{\bbU}_{s,q}(j))|)^{-\gamma}$. Since $\hat{\bbU}_{s,q}$ is an
asymptotically normal estimator for $\bbU_{s,q}$ it follows that
$(\sqrt{n}|\textrm{vec}(\hat{\bbU}_{s,q}(j))|)^{-\gamma}$
converges in distribution to a random variable of finite variance
as $n\rightarrow\infty$. Given that  $\lambda_{\rho,n}$ satisfies
the second limit in \eqref{Eq:Scaling_Law_lambda}, using the
previous two limits and Slutsky's theorem we have that the
quantity in \eqref{Eq:J_b_nz_J_b_0} converges in distribution to
\begin{equation}\label{Eq:Cost_Limit}
G_{b}(\tilde{\bbb})= \tilde{\bbb}_{{\cal
S}_{o}}^T[\bbD_{s,q}\otimes \bbI_{p}]_{{\cal
S}_{o}}\tilde{\bbb}_{{\cal S}_{o}}-2\tilde{\bbb}_{{\cal
S}_{o}}^T\left[\textrm{vec}(\bbU_{s,p-q}\bbU_{s,p-q}^T\bbSigma_s{\bbE}_{c}^T
-\bbU_{s,q}{\bbE}_{c}\bbSigma_{s}\bbU_{s,q}) \right]_{{\cal
S}_{o}}
\end{equation}
if $\tilde{\bbb}(j)=0$ for all $j\notin {\cal S}_{o}$; otherwise,
the limit is $\infty$. The notation  $[\bbM]_{{\cal S}_{o}}$ in
\eqref{Eq:Cost_Limit} denotes the submatrix of $\bbM$ whose row
and column indices are in ${\cal S}(\bbU_{s,q})$. The optimal
solution of \eqref{Eq:Cost_Limit} $\breve{\bbb}$ is given by
\begin{equation}\label{Eq:Breve_b}
\breve{\bbb}({\cal S}_{o})=\left([\bbD_{s,q}\otimes
\bbI_{p}]_{{\cal
S}_{o}}\right)^{-1}\left[\textrm{vec}(\bbU_{s,p-q}
\bbU_{s,p-q}^T\bbSigma_s{\bbE}_{c}^T
-\bbU_{s,q}{\bbE}_{c}\bbSigma_{s}\bbU_{s,q}) \right]_{{\cal
S}_{o}},\;\textrm{ and }\breve{\bbb}(\bar{{\cal
S}}_{o})=\mathbf{0}.
\end{equation}

Since  the cost in \eqref{Eq:Btaup1} is strictly convex wrt $\bbB$
and
$J_{b}(\tilde{\bbb})-J_{b}(\mathbf{0})\xrightarrow{d}G_{b}(\tilde{\bbb})$
as $n\rightarrow\infty$, one can readily apply the epi-convergence
results in \cite{Knight_Fu_Epiconvergence} to establish that
$\breve{\bbb}_n\xrightarrow{d}\breve{\bbb}$ as $n\rightarrow\infty$,
while $\breve{\bbb}$ corresponds to a zero-mean Gaussian random
vector. This establishes asymptotic normality of
$\hat{\bbB}_{\tau,n}$. An interesting thing to notice is that when
setting in \eqref{Eq:Btaup1} $\mu=0$ and
$\{\lambda_{\rho,n}=0\}_{\rho=1}^{q}$ (standard PCA approach) it
follows that the corresponding cost in \eqref{Eq:J_b_nz_J_b_0}
converges in distribution to the one in \eqref{Eq:Cost_Limit} with
${\cal S}_{o}=\{1,\ldots,p\}$. This result establishes that the
covariance of $\breve{\bbb}$ is equal to $[\bbSigma_{E_b}]_{{\cal
S}_o}$, where $\bbSigma_{E_b}$ is the limiting covariance matrix of
the estimation error when the standard PCA approach is employed.

Next, we prove that the probability of finding the correct support
converges to unity as $n\rightarrow\infty$. Letting $\hat{{\cal
S}}_{B,n}:={\cal S}(\hat{\bbB}_{\tau,n})$, we have to show that:
i) $\textrm{Pr}(\{i\in\hat{{\cal
S}}_{B,n}\})\xrightarrow{n\rightarrow\infty}1$ $\forall i\in{\cal
S}_{o}$; and ii) $\textrm{Pr} (\{i\in\hat{{\cal
S}}_{B,n}\})\xrightarrow{n\rightarrow\infty}0$ $\forall i\notin
{\cal S}_{o}$. The asymptotic normality of $\hat{\bbB}_{\tau,n}$
implies that $\hat{\bbB}_{\tau,n}\xrightarrow{d}\bbU_{s,q}$. Since
$\bbU_{s,q}$ is a constant matrix, it also holds that
$\hat{\bbB}_{\tau,n}\xrightarrow{p}\bbU_{s,q}^T$ and the first
part of the proof is established. Concerning the second part,
differentiate the cost in \eqref{Eq:Btaup1_reform} wrt $\bbb$, and
apply the first-order optimality conditions to obtain an equality
whose lhs and right hs (rhs) are normalized with $n$. It then
holds $\forall$ $j\in\{1,\ldots,p\}$ and $\rho\in\{1,\ldots,q\}$
for which $(\rho-1)p+j\in \hat{{\cal S}}_{B,n}$ that
\begin{align}\label{Eq:First_order_optimality}
&\frac{2\left[(\hat{\bbC}_{\tau,n}\bbS)\otimes
\bbI_{p}\right]_{(\rho-1)p+j:}^T
\left[((\hat{\bbC}_{\tau,n}\bbS)^T\otimes \bbI_{p})[\textrm{vec}
(\bbU_{s,q})-\hat{\bbb}_{\tau,n}] +\bbz_{q,n}-\sqrt{n^{-1}}
[({\bbE}_{c,n}\bbS)^T\otimes
\bbI_{p}]\textrm{vec}(\bbU_{s,q})\right]}{\sqrt{n}}
\nonumber\\
&\hspace{2cm}+\frac{2\mu(\hat{\bbb}_{\tau,n}((\rho-1)p+j)
-\hat{\bbc}_{\tau,n}^t((\rho-1)p+j))}{\sqrt{n}}
=\textrm{sgn}[\hat{\bbb}_{\tau,n}((\rho-1)p+j)]
\frac{\lambda_{\rho,n}\hat{w}_{j,\rho,n}}{\sqrt{n}}
\end{align}
where $[\bbM]_{j:}$ denotes the $j$th row of matrix $\bbM$ and
$\hat{\bbb}_{\tau,n}=\textrm{vec}(\hat{\bbB}_{\tau,n})$. The rhs
in \eqref{Eq:First_order_optimality} can be rewritten as
$\frac{\lambda_{\rho,n}}{\sqrt{n}}\frac{n^{\gamma/2}}
{|\sqrt{n}\hat{\bbU}_{s,q}(j,\rho)|^{\gamma}}$  and goes to
$\infty$ when $\{\lambda_{\rho,n}\}_{\rho=1}^{q}$ are selected
according to \eqref{Eq:Scaling_Law_lambda}. The second fraction at
the lhs of \eqref{Eq:First_order_optimality} converges to $0$ in
probability. Next, we show why the first and third terms in the
first fraction converge in distribution to a zero-mean Gaussian
variable with finite variance. For the first term this is true
because: i) $\hat{\bbC}_{\tau,n}\xrightarrow{p}{\bbU}_{s,q}^T$;
ii) $n^{-1}\bbS\bbS^T\xrightarrow{a.s.}\bbSigma_{s}$; and iii)
$\sqrt{n}[\textrm{vec}(\bbU_{s,q})-\hat{\bbb}_{\tau,n}]$ converges
in distribution to a zero-mean Gaussian random variable as shown
earlier. This is the case for
 the third term too since: i) $n^{-1}\bbS\bbS^T\xrightarrow{a.s.}\bbSigma_{s}$;
 ii) $\hat{\bbC}_{\tau,n}\xrightarrow{p}{\bbU}_{s,q}^T$;
 and iii) ${\bbE}_{c,n}$ converges
in distribution to  a zero-mean Gaussian random matrix. Finally,
the second term in the first fraction converges in probability to
a constant because
$n^{-1}\bbS\bbS^T\xrightarrow{a.s.}\bbSigma_{s}$ and
$\hat{\bbC}_{\tau,n}\xrightarrow{p}{\bbU}_{s,q}^T$.

Notice that the event $i=(\rho-1)p+j\in \hat{{\cal S}}_{B,n}$
implies equality \eqref{Eq:First_order_optimality}; hence,
$\forall i\notin{\cal S}_o$ $\textrm{Pr}(\{i\in\hat{{\cal
S}}_{B,n}\})\leq \textrm{Pr} (\{\textrm{eq.
\eqref{Eq:First_order_optimality} is true}\})$. However, as
$n\rightarrow\infty$ the probability of
\eqref{Eq:First_order_optimality} being satisfied goes to zero
since the lhs  converges to a Gaussian  variable and the rhs goes
to $\infty$; thus, $\forall i\notin{\cal S}_{o} $ it holds that
$\lim_{n\rightarrow\infty}\textrm{Pr}(\{i\in\hat{{\cal
S}}_{B,n}\})=0$.\myQED

\nocite{*}
\bibliographystyle{IEEEtranS}
\bibliography{biblio}


%
%

%

%
%
\end{document}